\documentclass[11pt]{article}
\pdfoutput=1 
\usepackage{jheppub}
\usepackage[utf8]{inputenc}

\usepackage{graphicx,ulem}
\usepackage{caption,epigraph}
\usepackage{subcaption}
\usepackage{hyperref}
\usepackage[T1]{fontenc}
\usepackage[export]{adjustbox}
\usepackage[table]{xcolor}
\usepackage{amssymb}
\usepackage{pifont}

\newcommand{\be}{\begin{equation}}
\newcommand{\ee}{\end{equation}}
\newcommand{\bea}{\begin{eqnarray}}
\newcommand{\eea}{\end{eqnarray}}

\usepackage{multirow, graphicx,amssymb,url,mathrsfs,amsmath}
\usepackage{amsxtra,amstext,latexsym,dsfont,amsfonts}
\usepackage{color,eucal}
\usepackage{float}
\usepackage{colortbl}
\usepackage{multirow}
\usepackage{tikz}
\usetikzlibrary{tikzmark}
\usetikzlibrary{arrows}
\usepackage{tabularx, array}
\newcolumntype{L}[1]{>{\raggedright\arraybackslash}p{#1}}
\newcolumntype{C}[1]{>{\centering\arraybackslash}p{#1}}
\newcolumntype{R}[1]{>{\raggedleft\arraybackslash}p{#1}}
\makeatother
\makeatletter
\newcommand{\myBig}{\bBigg@{1.75}}
\makeatother

\title{\Large Dynamical stability from quasi normal modes in 2nd, 1st and 0th order holographic superfluid phase transitions}

\author[a]{Zi-Qiang Zhao,}
\author[a,b]{Xing-Kun Zhang,}
\author[a,1]{Zhang-Yu Nie\note{Corresponding author.}}
\affiliation[a]{Kunming University of Science and Technology, Kunming 650500, China}
\affiliation[b]{College of Physics, Nanjing University of Aeronautics and Astronautics, Nanjing 211106, China}

\emailAdd{zhaoziqiang@stu.kust.edu.cn}
\emailAdd{zhangxk@nuaa.edu.cn}
\emailAdd{niezy@kust.edu.cn}

\abstract{We study a simple extension of the original Hartnoll, Herzog and Horowitz (HHH) holographic superfluid model with two nonlinear scalar self-interaction terms $\lambda |\psi|^4$ and $\tau |\psi|^6$ in the probe limit. Depending on the value of $\lambda$ and $\tau$, this setup allows us to realize a large spectrum of holographic phase transitions which are 2nd, 1st and 0th order as well as the ``cave of wind'' phase transition. We speculate the landscape pictures and explore the near equilibrium dynamics of the lowest quasinormal modes (QNMs) across the whole phase diagram at both zero and finite wave-vector. We find that the zero wave-vector results of QNMs correctly present the stability of the system under homogeneous perturbations and perfectly agree with the landscape analysis of homogeneous configurations in canonical ensemble. The zero wave-vector results also show that a 0th order phase transition cannot occur since it always corresponds to a global instability of the whole system. The finite wave-vector results show that under inhomogeneous perturbations, the unstable region is larger than that under only homogeneous perturbations, and the new boundary of instability match with the turning point of condensate curve in grand canonical ensemble, indicating a new explanation from the subsystem point of view. The additional unstable section also perfectly match the section with negative value of charge susceptibility.
}

\begin{document}
\maketitle
\flushbottom

\section{Introduction}
Phase transitions are ubiquitous in nature as they arise in very disparate systems from condensed matter to modern cosmology \cite{sachdev1999quantum}. The study of the stability and dynamics near phase transitions is an important and old subject in physics. Phase transitions are historically classified in terms of their \text{order}. The latter is defined as the order of the first discontinuous derivative of the {thermodynamic potential} with respect to a specific external parameter. A first order phase transition is, for example, the case in which the first derivative of the {thermodynamic potential} (the order parameter) is discontinuous. A second order phase transition corresponds to a situation in which the second derivative is discontinuous and the order parameter is continuous, and so on. Second order, or continuous, phase transitions are associated to spontaneous symmetry breaking in the common sense. Their dynamics is usually labelled as \textit{critical} and has been studied in many setups \cite{RevModPhys.49.435}. On the contrary, the dynamical behavior in discontinuous first order phase transitions is somehow more complicated and exhibits a more rich phenomenology including for example interesting phenomena such as \textit{bubbles nucleation}.

In the last decades, holography has emerged as a very convenient and efficient tool to investigate strongly coupled field theories at finite temperature \cite{zaanen2015holographic,Hartnoll:2016apf,Baggioli:2019rrs}. One of its most celebrated applications is the holographic realization of superfluid phase transitions~\cite{Hartnoll:2008vx}. The dynamical evolution and non-equilibrium physics of the holographic superfluid model have been investigated in several studies~ (see \cite{Adams:2013vsa,Chesler:2014gya,Guo:2018mip,Li:2019swh,Xia:2019eje,Yang:2019ibe,Adams:2012pj,Lan:2016cgl,Du:2014lwa,
Liu:2018crr,Zeng:2019yhi,Arean:2021tks,Ammon:2021pyz,Xia:2021pap} for a non exhaustive list) and even extended to supersolid phases \cite{Baggioli:2022aft} or in presence of momentum dissipation \cite{Ling:2014laa,Andrade:2014xca,Kim:2015dna,Baggioli:2015zoa}.

The holographic superfluid model has been enlarged to study more general order parameters, as well as the competition and coexistence of different orders \cite{Basu:2010fa,Musso:2013ija,Nie:2013sda,Donos:2013woa,Li:2017wbi,Nie:2015zia,Nie:2014qma,Amado:2013lia}. In this more complex scenarios, different types of phase transitions have been found including 1st order and 0th order phase transitions ~\cite{Cai:2013aca,Nie:2016pjt}. However, in all these cases, the bulk action needs an expanded matter content involving more bulk field components. As an immediate drawback, the numerical calculation costs more resources specially for the study of quasi normal modes or of the non-equilibrium and inhomogeneous dynamics. Moreover, it is convenient to study various kinds of phase transitions in a single unified setup. This is possible in a most simple setup in the probe limit with the nonlinear potential terms~\cite{Zhang:2021vwp} from the landscape point of view. 

Contrary to 2nd order and 1st order phase transitions, which are quite common, the physics related to 0th order phase transitions is quite mysterious and not completely understood yet. In a 0th order phase transition~\cite{Cai:2013aca,Nie:2016pjt}, the {thermodynamic potential} of the system is not continuous at the phase transition point. Moreover, a 0th order phase transition undergoes with the system jumping from a state with a lower value of the {thermodynamic potential} (e.g., free energy) to a solution with a higher one, which seems quite unnatural from a thermodynamic perspective. One obvious way to better understand this strange type of phase transition is to investigate the dynamical stability of the related solutions.

The stability of a system can be examined using different criteria and in particular one can distinguish between thermodynamic stability and dynamical stability. While the first one is derived from the equilibrium properties of the system, the second one necessitates the study of the perturbations away, and eventually far from the equilibrium configuration. In addition to that, thermodynamic stability can be further classified into global stability and local stability, amounts to the difference between a global or local minimum in the thermodynamic potential.
Either way, the relation between thermodynamic stability and dynamical stability and whether one implies the other and/or vice versa is not known in general terms. In the context of gravitational theories, the equivalence between thermodynamic and dynamical stability is known as \textit{the correlated stability conjecture} and it has been subject of many studies \cite{Gubser:2000ec,Gubser:2000mm,Buchel:2010wk,Buchel:2011ra,Friess:2005zp,Buchel:2005nt,Tian:2019fax}. The general validity of this conjecture is still a subject of debate.

On the other hand, at least in the linear response regime, the dynamical stability of the boundary system is a rather different concept which, in advantage of the gauge/gravity duality, can be directly verified using the so-called quasi normal modes of the classical bulk system. Quasinormal modes (QNMs) correspond to the poles of the retarded Green's functions and are characterized by their dispersion relation, i.e. the relation between their complex frequency $\omega$ and the real wave-vector $k$. Each QNM contributes to the response of the system with a factor $\propto \exp[-i \omega(k) t+ i \vec{k}\cdot\vec{x}]$ which is weighted by the corresponding residue. This implies that the condition of dynamical stability is equivalent to the requirement:
\begin{equation}
    \mathrm{Im}\left[\omega(k)\right]<0
\end{equation}
for all the modes in the spectrum. Whenever a QNM has a positive imaginary part at zero wave-vector, one does expect an instability under homogeneous perturbations. On the contrary, when the instability appears at finite wave-vector, one does expect an instability under inhomogeneous perturbations. Finally, the instability can be driven by hydrodynamic modes, which satisfy the condition $\omega(k=0)=0$, or by non-hydrodynamic ones which parametrize the more microscopic physical information and do not correspond to conserved quantities or Goldstone degrees of freedom.

QNMs of a bulk system can be computed using the linearized bulk gravitational dynamics and standard methods developed in the last decades \cite{PhysRevD.72.086009}. In the case of a 2nd order superfluid holographic phase transition, the QNMs in the probe limit have been studied in Ref.~\cite{Amado:2009ts}. The obtained results, defining the linear dynamical stability of the system, are fully consistent with the general wisdom about 2nd order phase transitions. The dynamical stability has been also investigated in more complicated holographic systems with competing orders, e.g., \cite{Du:2015zcb,Li:2020ayr}, and in presence of backreaction \cite{Arean:2021tks}.

First order phase transitions have been also subject of several studies recently because of their value both for condensed matter applications and cosmological problems \cite{https://doi.org/10.48550/arxiv.2209.12789,Attems:2017ezz,Attems:2018gou,Attems:2019yqn,Attems:2020qkg,Ares:2021nap,Li:2020ayr,Ares:2021ntv,Ares:2020lbt,Henriksson:2019ifu,Janik:2022wsx,Chen:2022tfy}. The dynamical stability of first order phase transitions has been studied using QNMs in \cite{Janik:2016btb,Janik:2015iry,Bellantuono:2019wbn}. In Ref.~\cite{Li:2020ayr}, the authors realized 1st order phase transitions in a model with two competing s-wave orders, and provided a snapshot of the QNMs at one point in the unstable region of the phase diagram. There, they found that the unstable mode in the homogeneous case, $k=0$, remains also unstable at finite wave-vector. A more detailed analysis in Ref.~\cite{Janik:2016btb} showed that in another holographic model presenting 1st order phase transitions, there is a small section on the spinodal branch near the tip point where the homogeneous modes are stable, but one hydrodynamic mode becomes unstable at finite wave-vector. Given these facts, it is natural to ask what is the boundary of the region that is stable at zero wave-factor but unstable at finite wave-vector and which is the most general picture. This is one of the open questions which we will address below.

More in general, in this work, we aim to generalize the previous analyses about the stability of holographic phase transitions to a much wider spectrum of cases, including 2nd order, 1st order and 0th order phase transitions, in a simple holographic model with a single scalar order parameter. Our main scope is to study the connection between the stability arguments derived from thermodynamics and the dynamical stability obtained from the QNMs analysis for a large class of phase transitions. Finally, we will try to extrapolate from all the examples some, possibly universal, conclusions which are expected to hold beyond the cases considered.\\

The manuscript is organized as follows. In Section \ref{sect:setup} we will introduce the holographic setup and the computational tools for both the analysis of the background solution and the perturbations around it; in Section \ref{sec2}, we will discuss the phase diagram, the potential landscape and the thermodynamic properties of the normal states and superfluid states for different types of phase transitions; in Section \ref{sec3}, we will study the dynamical stability by computing the QNMs at zero and finite wave-vector. Finally, in Section \ref{sect:conclusion}, we will conclude and present a short outlook as well as some perspectives for the future.

\section{Setup}  \label{sect:setup}
\subsection{Holographic model}
Let us consider a holographic s-wave superfluid model with nonlinear self-interaction terms in (3+1) dimensional asymptotic AdS spacetime. As a simple generalization of the original Hartnoll, Herzog and Horowitz (HHH) model \cite{Hartnoll:2008vx}, the action of the system is given by
\begin{align}
&S=\,S_{M}+S_{G}~,\quad
S_G=\,\frac{1}{2\kappa_g ^2}\int d^{4}x\sqrt{-g}\left(R-2\Lambda\right)~,\label{Lagg}\\
&S_M=\int d^{4}x\sqrt{-g}\left(-\frac{1}{4}F_{\mu\nu}F^{\mu\nu}
-D_{\mu}\psi^{\ast}D^{\mu}\psi  -m^{2}\psi^{\ast}\psi-\lambda(\psi^{\ast}\psi)^{2}-\tau(\psi^{\ast}\psi)^{3}\right)~.\label{Lagm}
\end{align}
Here, $D_{\mu}\psi=\nabla_{\mu}\psi-i q A_\mu\psi$ is the standard covariant derivative term and $F_{\mu\nu}=\nabla_{\mu}A_{\nu}-\nabla_{\nu}A_{\mu}$ is the Maxwell field strength. The bulk field $\psi$ is a complex scalar field charged under the bulk $U(1)$ gauge symmetry. We fix the value of the cosmological constant to $\Lambda=-3/L^2$ where $L$ is the AdS radius. The additional two terms appearing in the matter Lagrangian, Eq.\eqref{Lagm},  $\lambda(\psi^{\ast}\psi)^{2}$ and $\tau(\psi^{\ast}\psi)^{3}$, introduce nonlinear self-interactions for the bulk scalar field $\psi$, and play an important role in realizing holographic phase transitions beyond 2nd order.

In this paper, we work in the probe limit. Because of this approximation, the background geometry is fixed and parametrized by a $3+1$ dimensional black-brane metric element
\begin{align}
ds^{2}=-f(r)dt^{2}+\frac{1}{f(r)}dr^{2}+r^{2}dx^{2}+r^{2}dy^{2},
\end{align}
where the emblackening function $f(r)$ is given by
\begin{align}
f(r)=r^2\left(1-\frac{r_h^3}{r^3}\right)~.
\end{align}
Here, $r_{h}$ is the horizon radius at which the function $f(r)$ vanishes. The Hawking temperature corresponding to this black-brane solution is given by
\begin{align}
T= \frac{3 r_h}{4\pi L^2}.
\end{align}

For the background solution, we choose the following Ansatz for the matter bulk fields
\begin{align}\label{ansatz}
\psi=\psi(r)~, \quad A_{t}=\phi(r)~,
\end{align}
with a real valued function $\psi(r)$ and the other components of the gauge field $A_\mu$ set to zero. This choice corresponds to a dual field theory at finite temperature and finite charge density in presence of a charged scalar operator, dual to the bulk field $\psi$, which might acquire a non-trivial expectation value $\langle \mathcal{O}\rangle$.
Using this ansatz, the equations of motion for the matter fields are
\begin{align}
&\phi''+\frac{2}{r}\phi'-2\frac{q^{2}\psi^{2}}{f}\phi=0,\label{EqPhi}\\
&\psi''+\left(\frac{f'}{f}+\frac{2}{r}\right)\psi'
+\left(\frac{q^{2}\phi^{2}}{f^{2}}-\frac{m^{2}}{f}\right)\psi-\frac{2\lambda}{f}\psi^{3}
-\frac{3\tau}{f}\psi^{5}=0.\label{EqPsis}
\end{align}
In the rest of the paper, we focus on the effects of the two nonlinear coefficients $\lambda$ and $\tau$. Therefore, for simplicity, we set $m^2=-2$, $q=1$ and $r_h=L=1$.

To solve the equations of motion, it is necessary to specify boundary conditions both at the horizon and on the boundary. The asymptotic expansions of the two bulk fields at the horizon are
\begin{align}
&\phi(r)=\phi_{1}(r-r_{h})+\mathcal{O}((r-r_{h})^{2})~,\\
&\psi(r)=\psi_{0}+\psi_{1}(r-r_{h})+\mathcal{O}(r-r_{h})~.
\end{align}
The expansions near the AdS boundary are
\begin{align}
\phi(r)=\mu-\frac{\rho}{r}+...~,\qquad \psi=\frac{{\psi^{(1)}}}{r}+\frac{{\psi^{(2)}}}{r^2}...~,
\end{align}
where $\mu$ and $\rho$ are respectively the chemical potential and the charge density of the dual field theory. For the bulk field $\psi$, we choose the standard quantization scheme and set ${\psi^{(1)}}=0$ as the source free boundary condition. The system of equations above has two kinds of solutions. The first has a vanishing scalar field and it corresponds to the normal state with unbroken U(1) symmetry in the dual field theory. On the contrary, the second, with finite value for the scalar field, corresponds to a superfluid state with spontaneously broken U(1) symmetry in which the charged operator dual to the bulk field $\psi$ acquires a non-trivial vacuum expectation value $\langle \mathcal{O}\rangle =\psi^{(2)}\neq 0$. 

In order to study the global thermodynamic stability of these two solutions, we need to compare the thermodynamic potential of the two different bulk solutions. Here we work in the canonical ensemble, in which the charge density $\rho$ is fixed, and the corresponding thermodynamic potential is the Gibbs free energy, which can be obtained from the Legendre-transformed Euclidean on-shell action (see for example footnote 13 in \cite{Hartnoll:2009sz}). As a consequence of the probe limit approximation, only the matter contribution
\begin{align}
	G=\frac{V_2}{T}\left[\frac{\mu\rho}{2L^2}+\int_{r_h}^{\infty}\left(\frac{q^2 r^2 \phi^2 \psi^2}{f}-r^2 \lambda\psi^4-2r^2\tau\psi^6\right)dr\right],
\end{align}
need to be considered to compare the values of differernt solutions. Here, $V_2$, is just the volume of the spatial boundary manifold.

\subsection{Fluctuations and numerical scheme}
Given a specific background solution, defined by the profiles of $\phi(r)$ and $\psi(r)$, we are then interested in computing linear perturbations around it. This will allow us to compute the lowest quasi normal modes in the dual field theory. The latter correspond to the poles of the field theory retarded Green's functions which can be obtained numerically in holography following standard methods \cite{PhysRevD.72.086009}. In order to do that, we consider the following fluctuations for the bulk fields
\begin{align}
\qquad
\delta \Psi=\tilde{\sigma}(r,t,x)+i\tilde{\eta}(r,t,x)~,\qquad \delta A_\mu=\tilde{a}_\mu(r,t,x)~.
\end{align}
By expanding the equations of motion at linear order in the perturbations, and going to Fourier space using the prescription $e^{-i(\omega t-k x)}$, the equations for the fluctuations are finally given by
\begin{align}
f\eta''+\left(f'+\frac{2f}{r}\right)\eta'+\myBig(\frac{\phi^2}{f}+\frac{\omega^2}{f}-\frac{k^2}{r^2}-m^2-2\lambda\psi^2&&\nonumber\\
-3\tau\psi^4\myBig)\eta-\frac{2i\omega\phi}{f}\sigma-\frac{i\omega\psi}{f}a_t-\frac{ik\psi}{r^2}a_x & =0~,&
\\ f\sigma''+\left(f'+\frac{2f}{r}\right)\sigma'+\myBig(\frac{\phi^2}{f}+\frac{\omega^2}{f}-m^2-\frac{k^2}{r^2}-6\lambda\psi^2&&\nonumber\\
-15\tau\psi^4\myBig)\sigma+\frac{2\phi\psi}{f}a_t+\frac{2i\omega\phi}{f}\eta&=0~,&
\\
f{a''_t}+\frac{2f}{r}a'_t-\left(\frac{k^2}{r^2}+2\psi^2\right)a_t-\frac{\omega{k}}{r^2}a_x-2i\omega\psi\eta
-4\psi\phi\sigma&=0~,&&
\\
\noindent{f}a''_x+f'a'_x+\left(\frac{\omega^2}{f}-2\psi^2\right)a_x+\frac{\omega{k}}{f}a_t+2ik\psi\eta&=0~,&
\end{align}
together with the constraint
\begin{align}
2i\eta\psi'-2i\psi\eta'-\frac{ka'_x}{r^2}-\frac{\omega{a'_t}}{f}=0~.
\end{align}

In order to calculate the quasi normal modes, we impose ingoing boundary condition at the horizon $r=r_h=1$,
\begin{align}
&\eta(r)=(r-1)^\xi(\eta^{(0)}+\eta^{(1)}(r-1)+...)~,\\
&\sigma(r)=(r-1)^\xi(\sigma^{(0)}+\sigma^{(1)}(r-1)+...)~,\\
&a_t(r)=(r-1)^{\xi+1}(a^{(0)}_t+a^{(1)}_t(r-1)+...)~,\\
&a_x(r)=(r-1)^{\xi}(a^{(0)}_x+a^{(1)}_x(r-1)+...)~,
\end{align}
with $\xi=-i\omega/3$. Because of the choice of ingoing boundary conditions
and the constraint equation, we have three linearly independent solutions characterized by the horizon coefficients. An additional fourth solution is given by the pure gauge solution
\begin{align}
\eta^{IV}=i\lambda\psi~,\sigma^{IV}=0~,a^{IV}_t=\lambda\omega~,a^{IV}_x=-\lambda k~,
\end{align}
with $\lambda$ an arbitrary gauge parameter which can be consistently set to unity.

On the contrary, close to the UV boundary located at $r=\infty$, the fluctuations have the following asymptotic behavior:
\begin{align}
&\eta(r)=\eta^{(L)}\left(1+\dots\right)\,\frac{1}{r}+\eta^{(S)}\left(1+\dots\right)\,\frac{1}{r^2}~,\\
&\sigma(r)=\sigma^{(L)}\left(1+\dots\right)\,\frac{1}{r}+\sigma^{(S)}\left(1+\dots\right)\,\frac{1}{r^2}~,\\
&a_t(r)=a_t^{(L)}\left(1+\dots\right)\,+a_t^{(S)}\left(1+\dots\right)\,\frac{1}{r}~,\\
&a_x(r)=a_x^{(L)}\left(1+\dots\right)\,+a_x^{(S)}\left(1+\dots\right)\,\frac{1}{r}~\,
\end{align}
where the labels $(L),(S)$ stand for leading and subleading. For all the fluctuations, we imposed standard boundary conditions $\eta^{(L)}=\sigma^{(L)}=a_t^{(L)}=a_x^{(L)}=0$. We should stress that, in general case involving time evolution, including the study of QNMs, the boundary condition for fix the chemical potential is rather subtle. However, the charge conservation still holds unless we impose conditions that explicitly break the U(1) symmetry. Therefore in our choice of boundary conditions, the QNMs indicate the dynamical stability of the boundary system in canonical ensemble where the total charge of the system is fixed.

In order to find numerically the quasi normal modes, we use the determinant method introduced in Ref.~\cite{Amado:2009ts}. Before continuing, let us emphasize that this model presents another advantage. Since the main modifications to construct more general phase transitions are nonlinear terms in the bulk field $\psi$, the position of the critical point $\rho=\rho_c$, where $\psi$ is small, is not modified with respect to the original HHH model. Moreover, the QNMs in the normal phase, with $\psi=0$, are also left unchanged and correspond to those found in Ref.~\cite{Amado:2009ts}.

\section{Phase transitions, thermodynamics and potential landscape}\label{sec2}
In the s-wave holographic superfluid model in the probe limit without the nonlinear terms, $\lambda=\tau=0$, the superfluid phase transition is always second order. In Ref.~\cite{Herzog:2010vz}, it was shown that when the coefficient $\lambda$ of the fourth power potential term acquires a lower enough negative value, the superfluid phase transition becomes either a first order phase transition or a runaway pathology. Ref.~\cite{Zhang:2021vwp} showed that tuning the value of the quartic coefficient $\lambda$ has non trivial effects on the dynamics of phase transitions involving multi-condensates as well. Inspired by these results~\cite{Herzog:2010vz,Zhang:2021vwp}, we expect to realize new kinds of phase transitions beyond second order by dialing the nonlinear $\lambda$ and $\tau$ terms.

In this section, we analyze in detail the phase diagram and the thermodynamic properties of the different kinds of phase transitions. We also illustrate how to rationalize these properties using the so-called landscape analysis.
\subsection{Second-order phase transition}
\begin{figure}
	\centering
	\includegraphics[width=0.33\columnwidth]{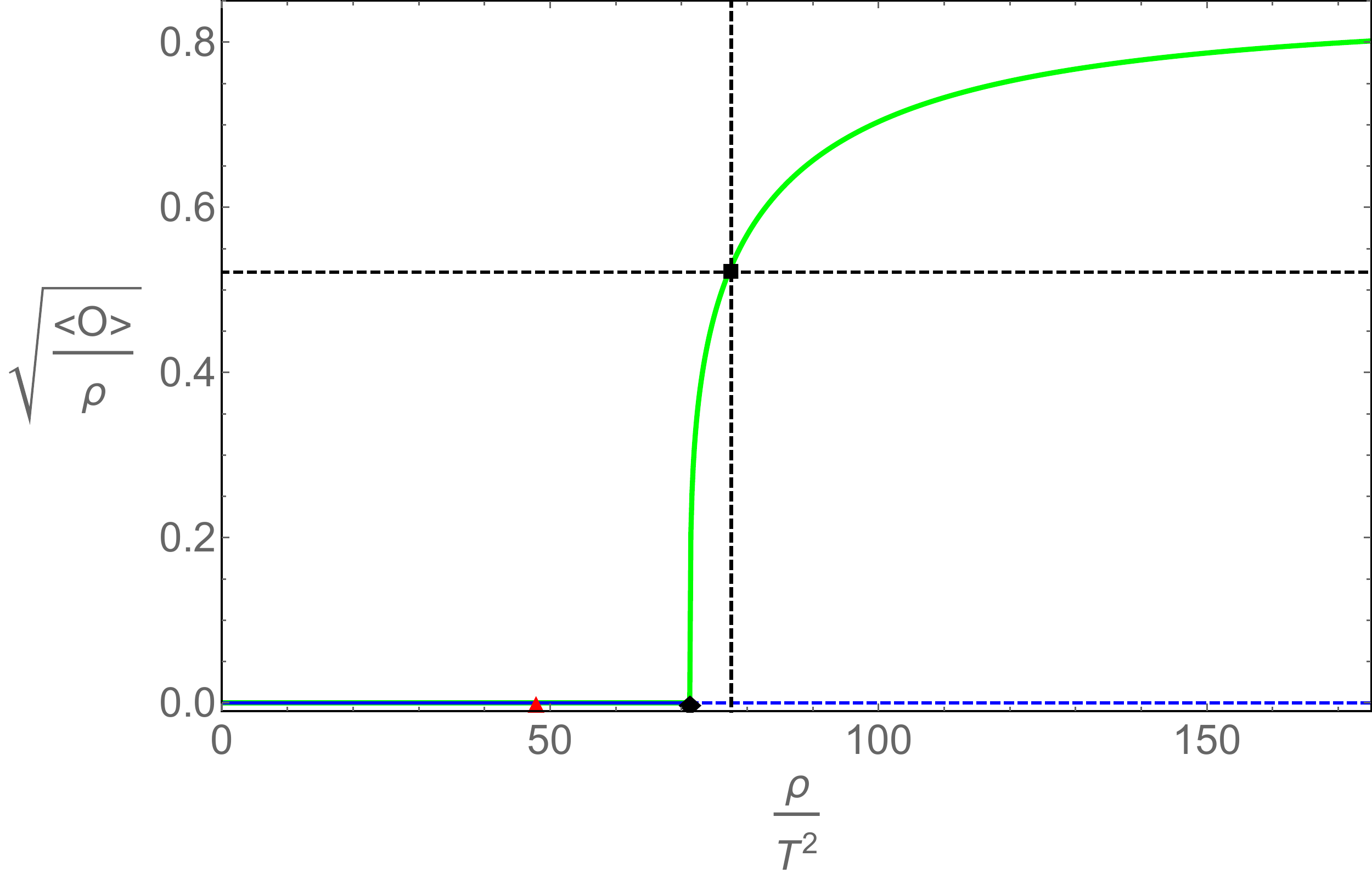}~~~~
    \includegraphics[width=0.33\columnwidth]{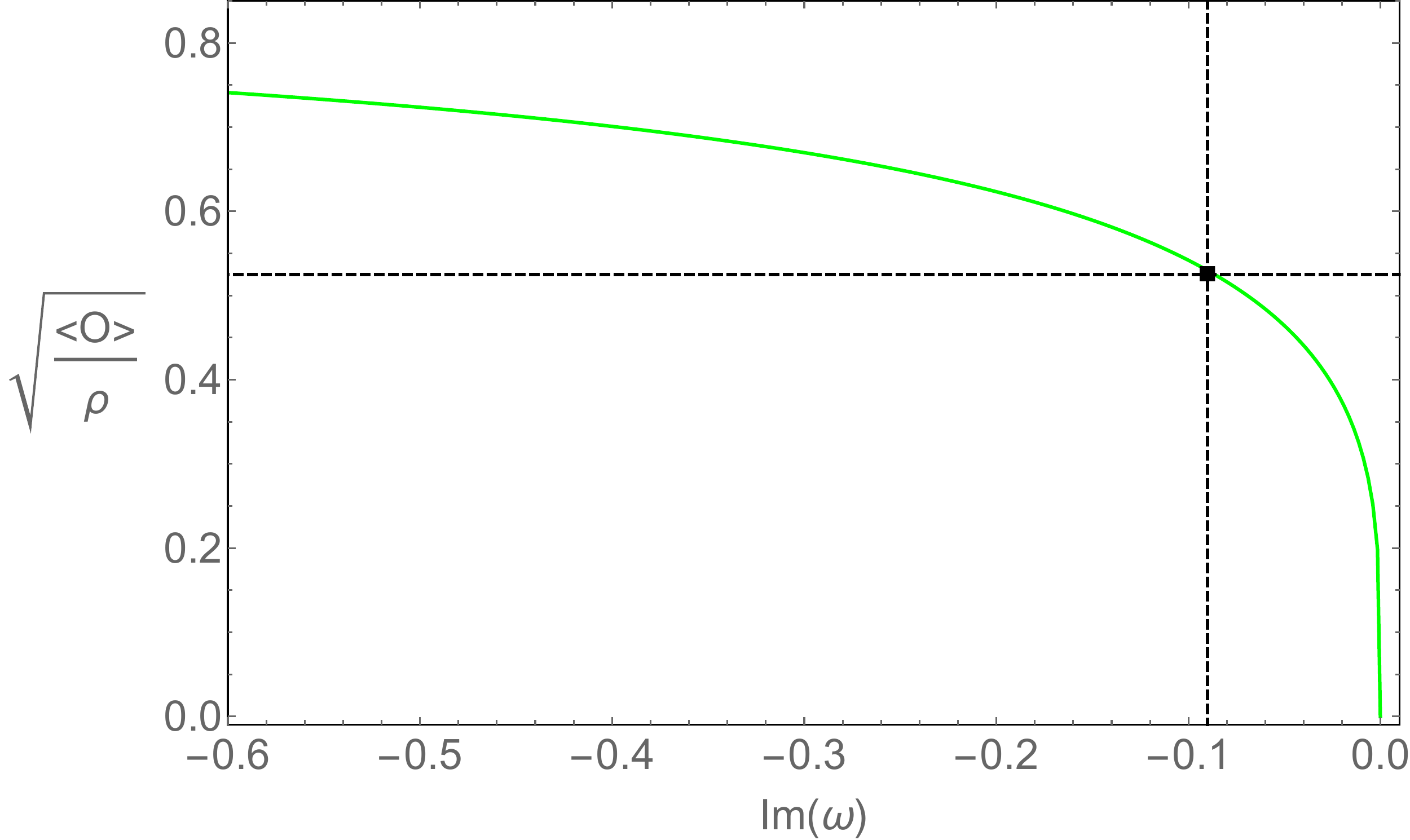}~~~~
    \includegraphics[width=0.33\columnwidth]{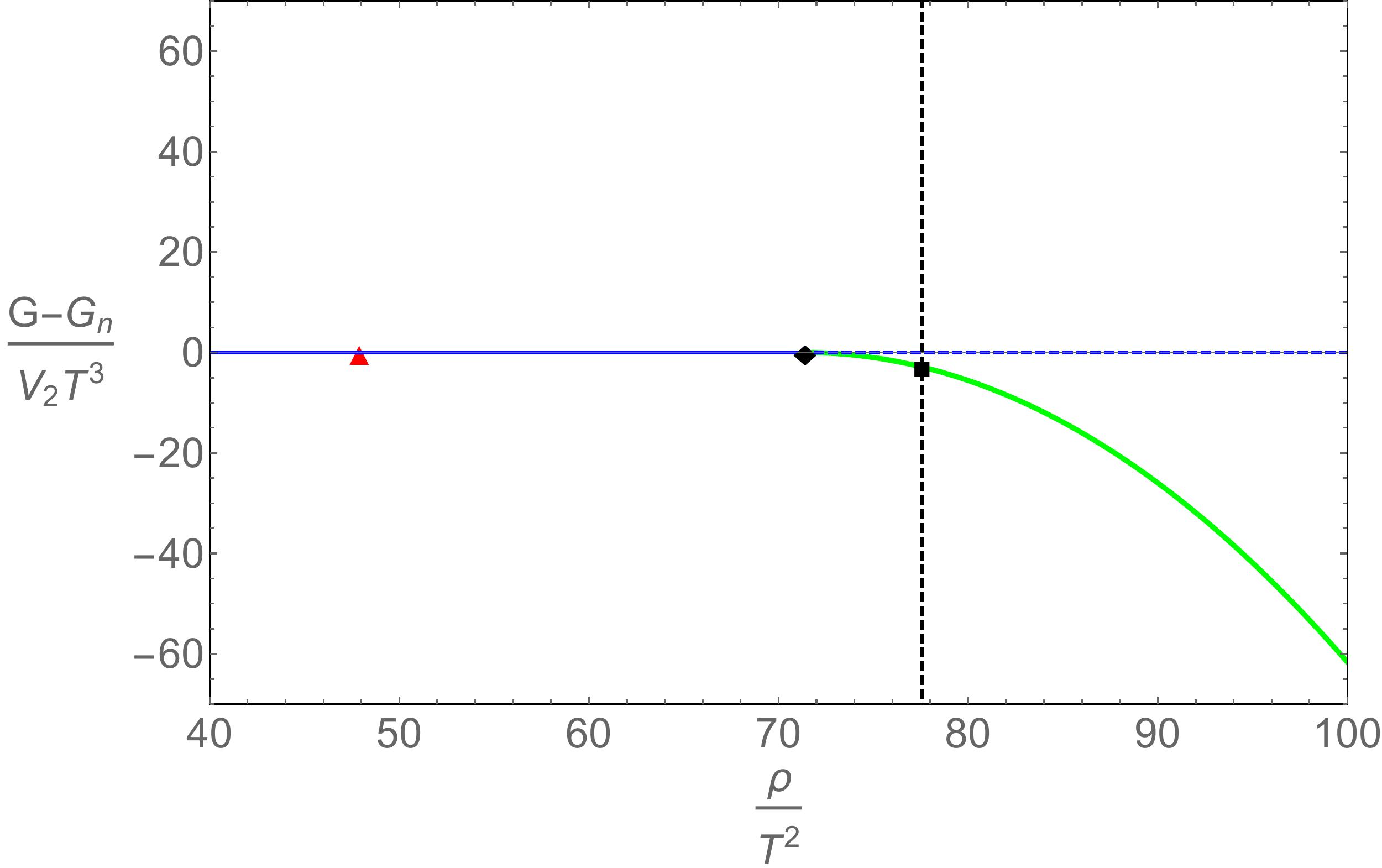}
	\caption{A typical 2nd order phase transition. \textbf{Left: } The normalized scalar condensate as a function of the dimensionless charge density. \textbf{Middle: } The imaginary part of the amplitude mode of the superfluid phase. \textbf{Right: } The {Gibbs free energy} difference between the normal phase and the superfluid phase. All figures are for $\lambda=\tau=0$ and $m^2=-2$, $q=1$. The black rhombus indicates the position of the critical point. The square and the triangle {as well as the dashed black lines are used to mark the same position in different plots}.}\label{condensate_free_2nd}
\end{figure}

We start with the simplest case in which the nonlinear terms in the action are switched off, $\lambda=\tau=0$. In this limit, we are considering the original holographic superfluid model of Ref.\cite{Hartnoll:2008vx}. Because we only consider the case $m^2=-2$, $q=1$, we obtain a simple 2nd order phase transition within the standard quantization scheme. We show the behavior of the dimensionless scalar condensate $\langle \mathcal{O} \rangle$ in function of the normalized charge density in the left panel of Figure.~\ref{condensate_free_2nd}. We further calculate the relative value of {Gibbs free energy} density with respect to that of the normal phase in the canonical ensemble. The results are presented in the right panel of Figure.~\ref{condensate_free_2nd}.

From there, we can see that the system undergoes a second order phase transition at the critical point with $\rho_c/T^2\approx 71.3$, which is a universal value in this study. The phase transition is of mean-field type, $\langle \mathcal{O} \rangle \propto \left(\rho-\rho_c\right)^{1/2}$ near the critical point, as expected \cite{Maeda:2009wv}. The first order derivative of the grand potential density is continuous at the critical point. The superfluid phase becomes thermodynamically favorable above the critical value for the charge density, $\rho_c$.

\subsection{Zeroth-order phase transition}
\begin{figure}
	\includegraphics[width=0.3\columnwidth]{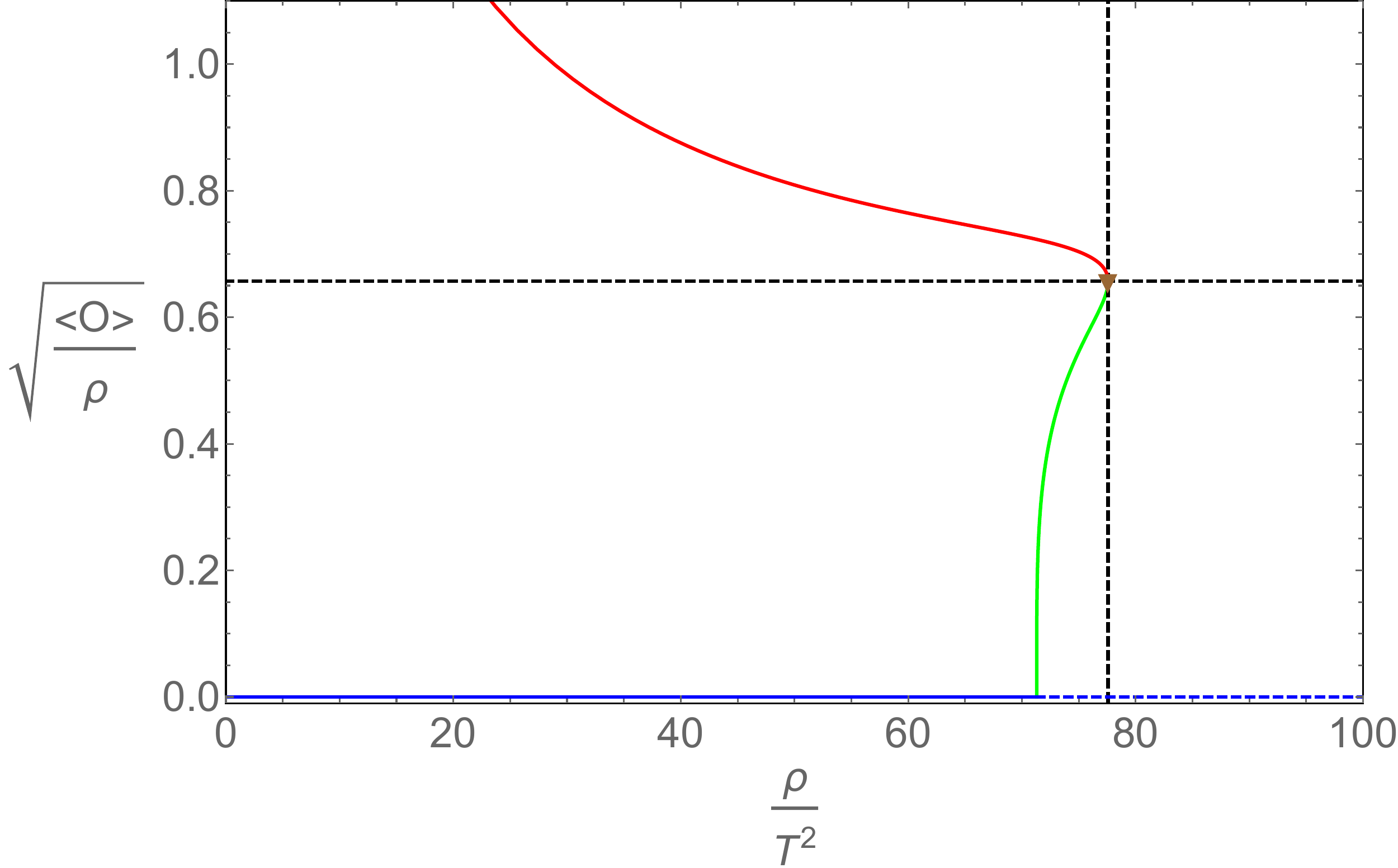}~~~~
    \includegraphics[width=0.3\columnwidth]{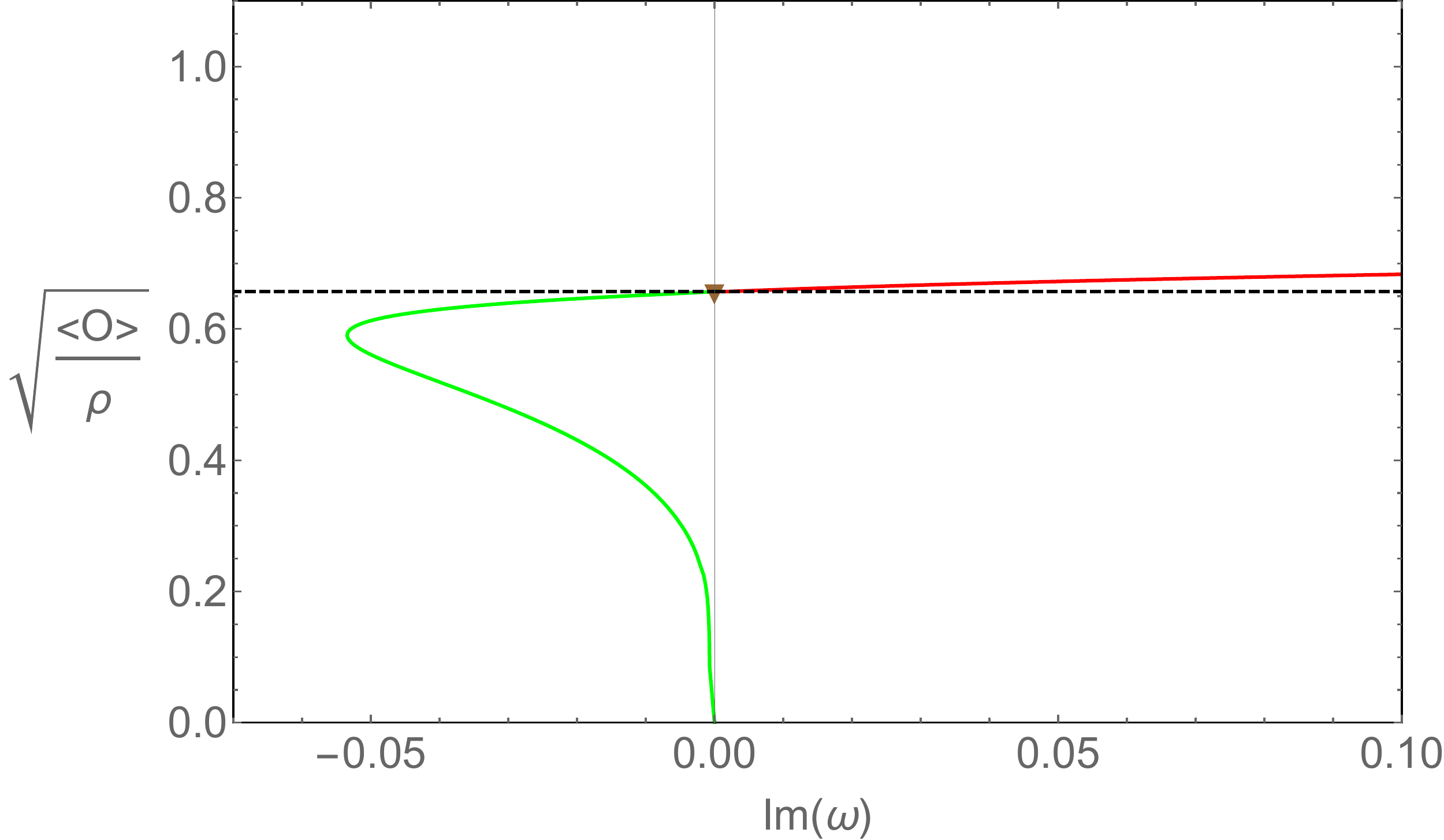}~~~~
    \includegraphics[width=0.3\columnwidth]{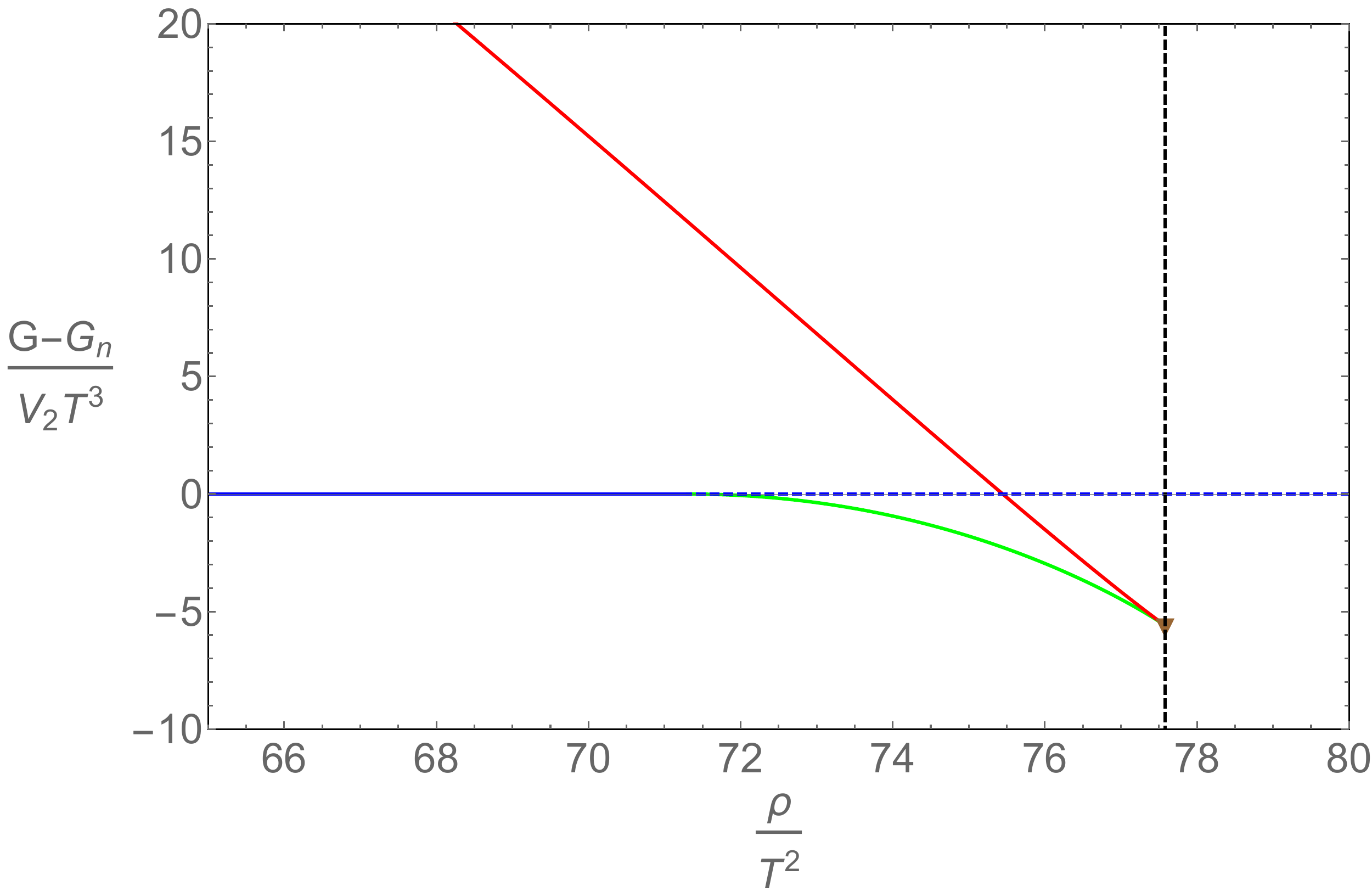}
	\caption{A typical zeroth order phase transition. \textbf{Left: } The normalized scalar condensate as a function of the dimensionless charge density. \textbf{Middle: } The imaginary part of the amplitude mode of the superfluid phase. \textbf{Right: } the {Gibbs free energy} difference between the normal phase and the superfluid phase. All figures are for $\lambda=-0.8,\tau=0$ and $m^2=-2$, $q=1$. The brown triangle indicates the position of the turning point at $\rho=\rho_t$.
	}\label{condensate_free_0th}
\end{figure}

A typical zeroth order phase transition can be simply realized by taking a finite and negative value of $\lambda$. As a benchmark example, here we choose $\lambda=-0.8$. The scalar condensate and the grand potential density for this case are shown in the left and right panels in Figure.~\ref{condensate_free_0th}. We can see that there is at first a second order phase transition with critical point $\rho=\rho_c$. Above this critical point, the scalar condensate $\langle \mathcal{O} \rangle$ first increases with the charge density. Nevertheless, beyond a specific value of the charge density, which we will call the \textit{turning point} with $\rho=\rho_t$, the superfluid phase does not exist anymore and the condensate turns back and continues growing in the opposite direction as indicate by the red branch in the left panel of Figure.~\ref{condensate_free_0th}. For our choice of parameters, the turning point is located at $\rho/T^2=\rho_t/T^2\approx 77.58$. Notice that, in the interval $\rho_c<\rho<\rho_t$, there are two different superfluid branches indicated respectively in green and red in the left panel of Figure.~\ref{condensate_free_0th}. Following the value of the scalar condensate along the two different branches, we will call the green one the \textit{bottom branch} and the red one the \textit{top branch}. By looking at the {Gibbs free energy} in the right panel of Figure.~\ref{condensate_free_0th}, we conclude that the Gibbs free energy density of the top branch is always larger than that of the bottom one. This implies that the top branch is not thermodynamically stable and can be considered only as an excited state. As we will see from the landscape analysis, the top branch corresponds to saddle points in the potential landscape which are not a minimum of the Gibbs free energy. In summary, in this situation, by increasing the charge density $\rho$, the system first undergoes one second order phase transition at $\rho=\rho_c$ and then another phase transition at $\rho=\rho_t>\rho_c$, after which no superfluid phase exists anymore. At this phase transition point $\rho=\rho_t$, the Gibbs free energy of the system jumps from a lower value to a higher value and it is not continuous. This type of behavior is referred to as the zeroth order phase transition~\cite{Cai:2013aca}.

{As reported in Ref.~\cite{Herzog:2010vz}, the condensate goes to the opposite direction at the critical point if the value of $\lambda$ is lower than some typical value. For our setup in the probe limit with 3+1 dimensional planar AdS$^4$ black brane back ground, this special value is $\lambda_s=-1.949$. We further confirmed that when $\lambda<\lambda_s$, the Gibbs free energy of the superfluid solution is always larger than that of the normal solution. Therefore 2nd order phase transition from normal phase to superfluid phase disappear, and the 0th order phase transition, in which the system jump back to normal phase from the superfluid phase with lower value of Gibbs free energy, also disappear. We call this situation the ``no stable superfluid phase''(NSSP) case for better references in the rest of this paper.
}

\subsection{First-order phase transition}
A system with a negative value of the quartic parameter $\lambda$ has a {Gibbs free energy} which is unbounded from below and it is therefore problematic in the limit of $\psi \rightarrow \infty$. To solve this issue, it is sufficient to introduce new positive terms in the potential which are dominant for $\psi\rightarrow \infty$. In other words, a model with a negative $\lambda$ parameter can be understood as an effective theory valid only in the small $\psi$ regime which must be modified away from it in order to avoid runaway pathologies. The simplest way is to consider a higher power term $\tau \psi^6$ with positive-valued coefficient $\tau$. After this higher-order term is included, the thermodynamic potential becomes bounded from below no matter how negative the quartic $\lambda$ parameter is.

However, the $\tau$ term significantly changes the solution only in the region where the condensate is large. If we add a positive $\tau$ term from the 0th order case with the value $\lambda=-0.8$, the condensate still grows with a increasing value of $\rho$ near the critical point, and we finally get a more complicated ``cave of wind'' type phase transition (or a 2nd order phase transition when $\tau$ is large enough). In order to get a standard 1st order phase transition from the normal phase to the superfluid phase, we need the condensate to grow in an opposite direction at the critical point $\rho=\rho_c$. In the above study on the 0th order phase transitions, We find that with $\lambda<\lambda_s=-1.949$, the condensate grows leftwards from the critical point, which is not affected by the value of $\tau$. Therefore, for simplicity, we choose $\lambda=-2<\lambda_s$ and further choose $\tau=0.8$ to study the 1st order phase transition.

\begin{figure}
	\includegraphics[width=0.3\columnwidth]{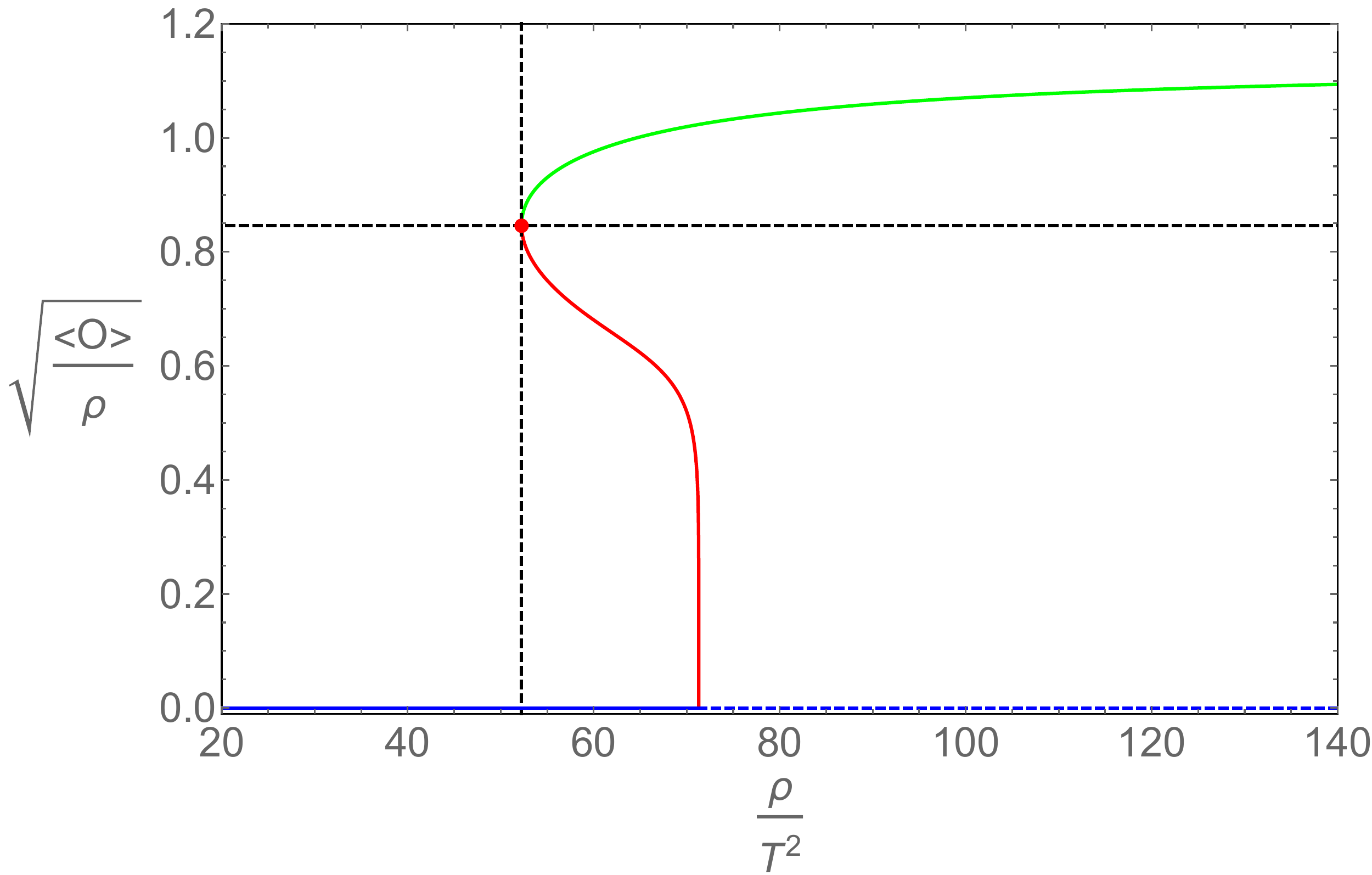}~~~~
    \includegraphics[width=0.3\columnwidth]{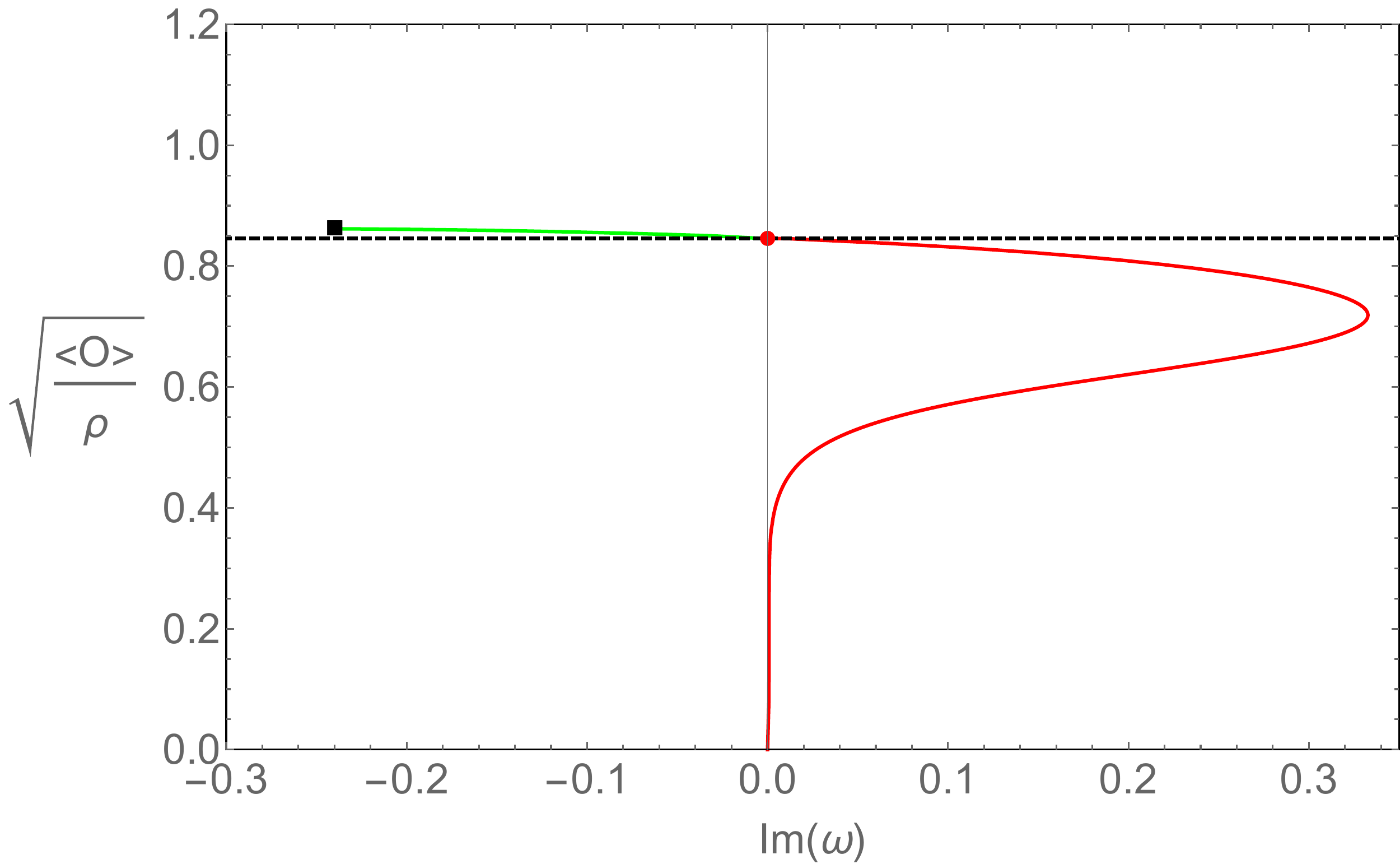}~~~~
    \includegraphics[width=0.3\columnwidth]{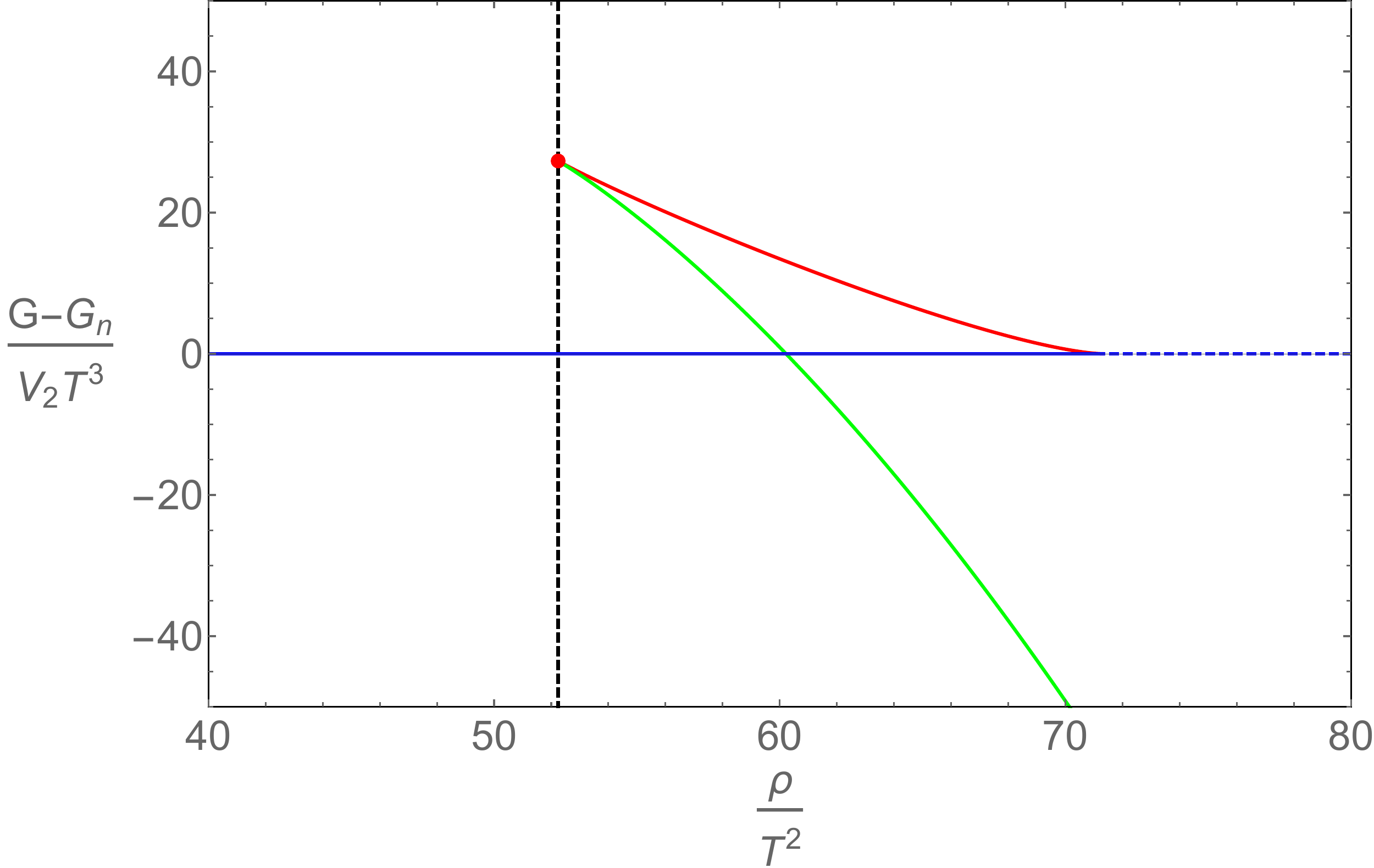}
	\caption{A typical first order phase transition. \textbf{Left: } The normalized scalar condensate as a function of the dimensionless charge density. \textbf{Middle: } The imaginary part of the amplitude mode of the superfluid phase. \textbf{Right: } the {Gibbs free energy} difference between the normal phase and the superfluid phase. All figures are for $\lambda=-2,\tau=0.8$ and $m^2=-2$, $q=1$. The red dot indicates the turning point. The black square denotes the position where the amplitude mode collide with another mode on the imaginary axes.
	}\label{condensate_free_1st}
\end{figure}

For convenience, we keep the concept of ``critical point'' to denote the point where the condensate solution is connected to the normal solution in the study of the 1st order phase transition. Near this critical point with small $\psi$, we still get the branch of saddle point solutions which move leftwards when the condensate is increasing (red line in Figure \ref{condensate_free_1st}). In addition to that, because of the positive value of $\tau$, there must be another minimum between the saddle points and the asymptotic solutions at $\psi\rightarrow\infty$ with positive infinite {Gibbs free energy}. These new stable minima form a new branch in the phase diagram which is now going rightwards with increasing condensate. This new branch is indicated with green color in Figure \ref{condensate_free_1st}. In the left plot of Figure \ref{condensate_free_1st}, we can see that the condensate first grows leftwards up to a specific point, labelled as the turning point, and then changes direction at the moment in which the new branch of (meta-)stable minima appears. It is expected that the phase transition occurs at the intersection point between the green and blue {Gibbs free energy} curves. As evident from the right panel of Figure \ref{condensate_free_1st}, the {Gibbs free energy} has now a discontinuous first derivative at the phase transition point, as expected.

\subsection{Cave of wind phase transition}
{Let us discuss a more exotic type of phase transition known as the \textit{cave of wind} phase transition. This is not a novel class of phase transitions beyond the 0th, 1st and 2nd order ones. It is in fact a first order phase transition between two section of superfluid phases with the same order parameter. We study this kind of phase transition to help understand how we rescue a problematic 0th order phase transition to this exotic type later.}

\begin{figure}
	\includegraphics[width=0.3\columnwidth]{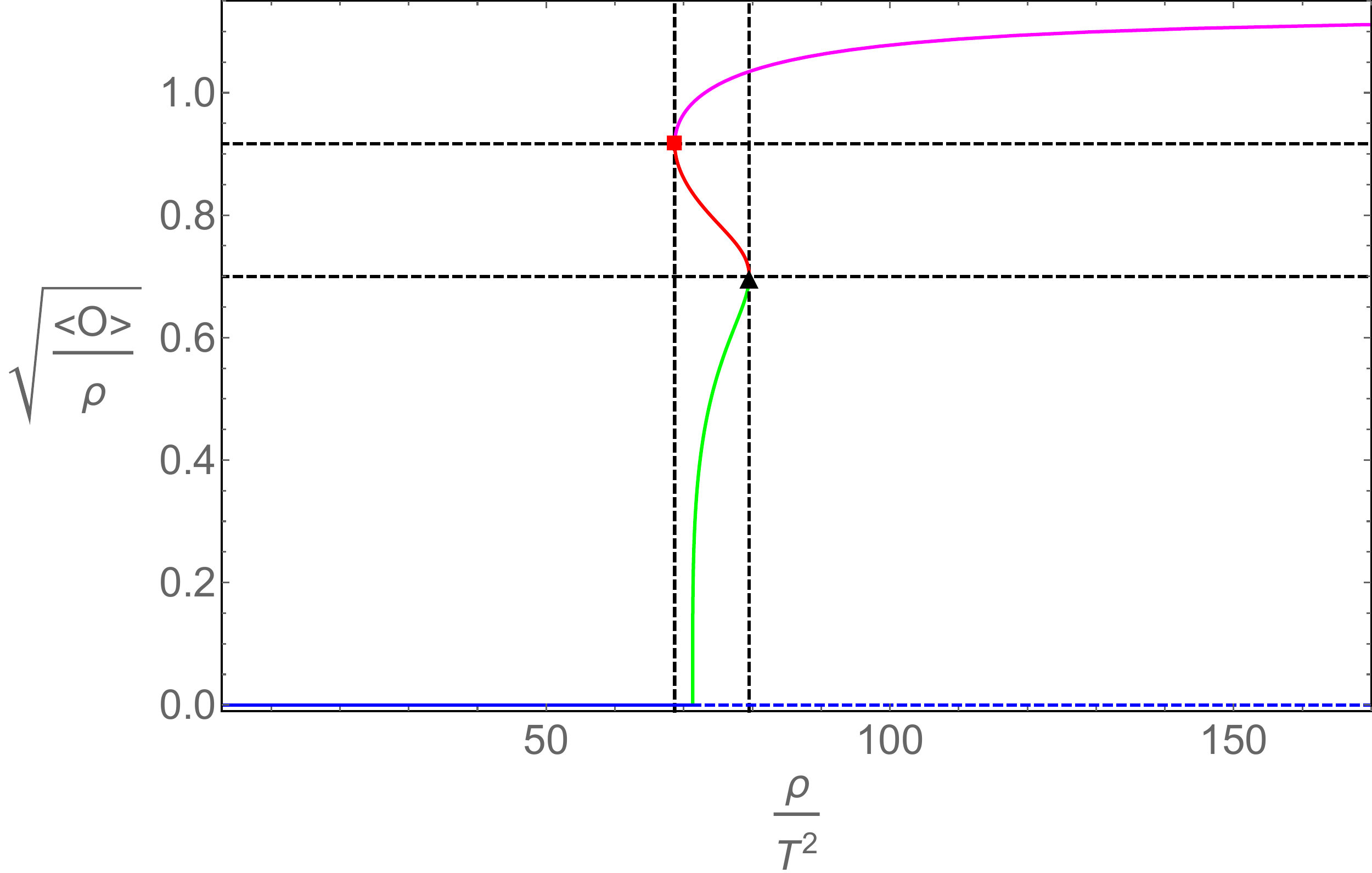}~~~~
    \includegraphics[width=0.3\columnwidth]{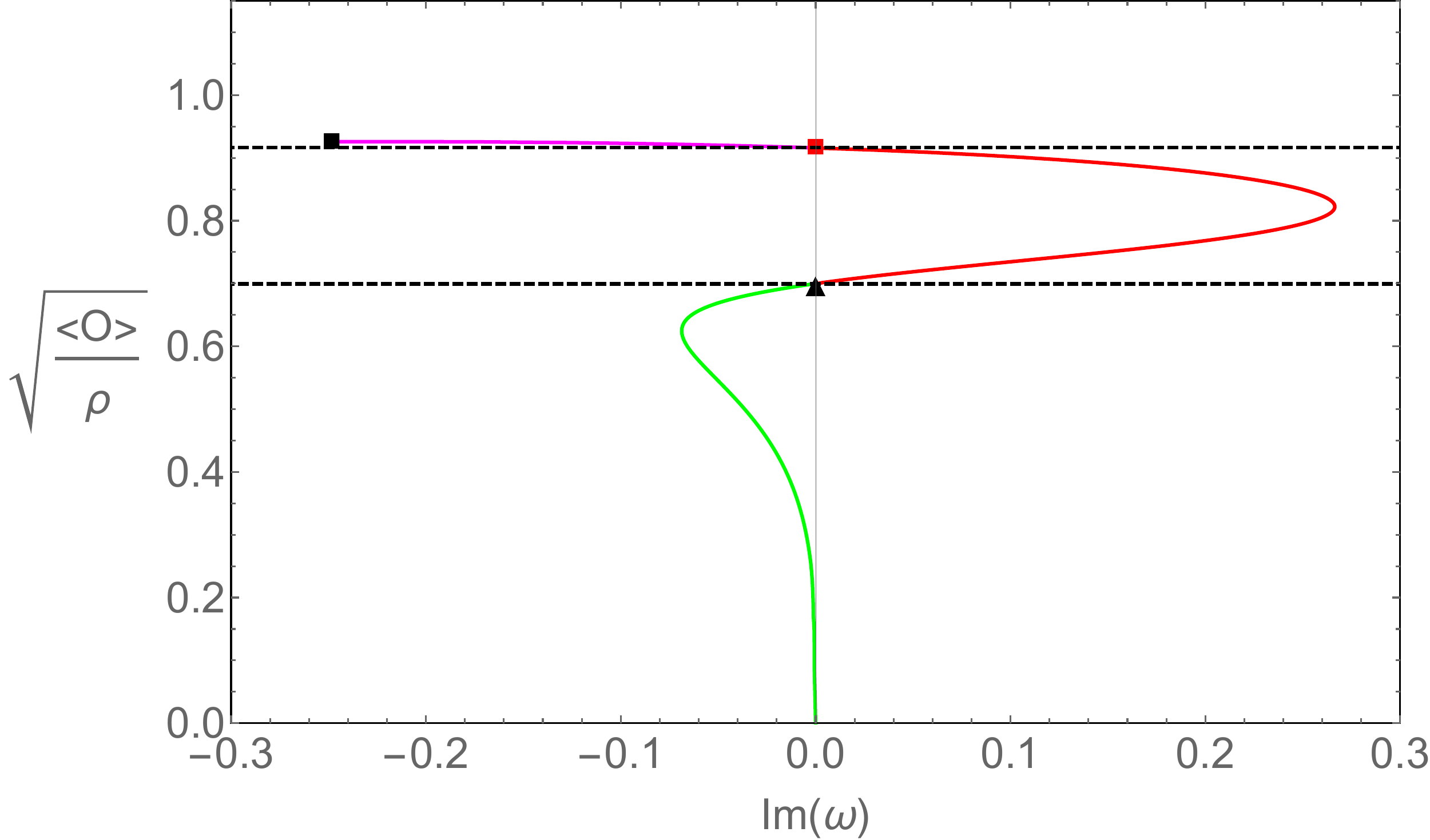}~~~~
    \includegraphics[width=0.3\columnwidth]{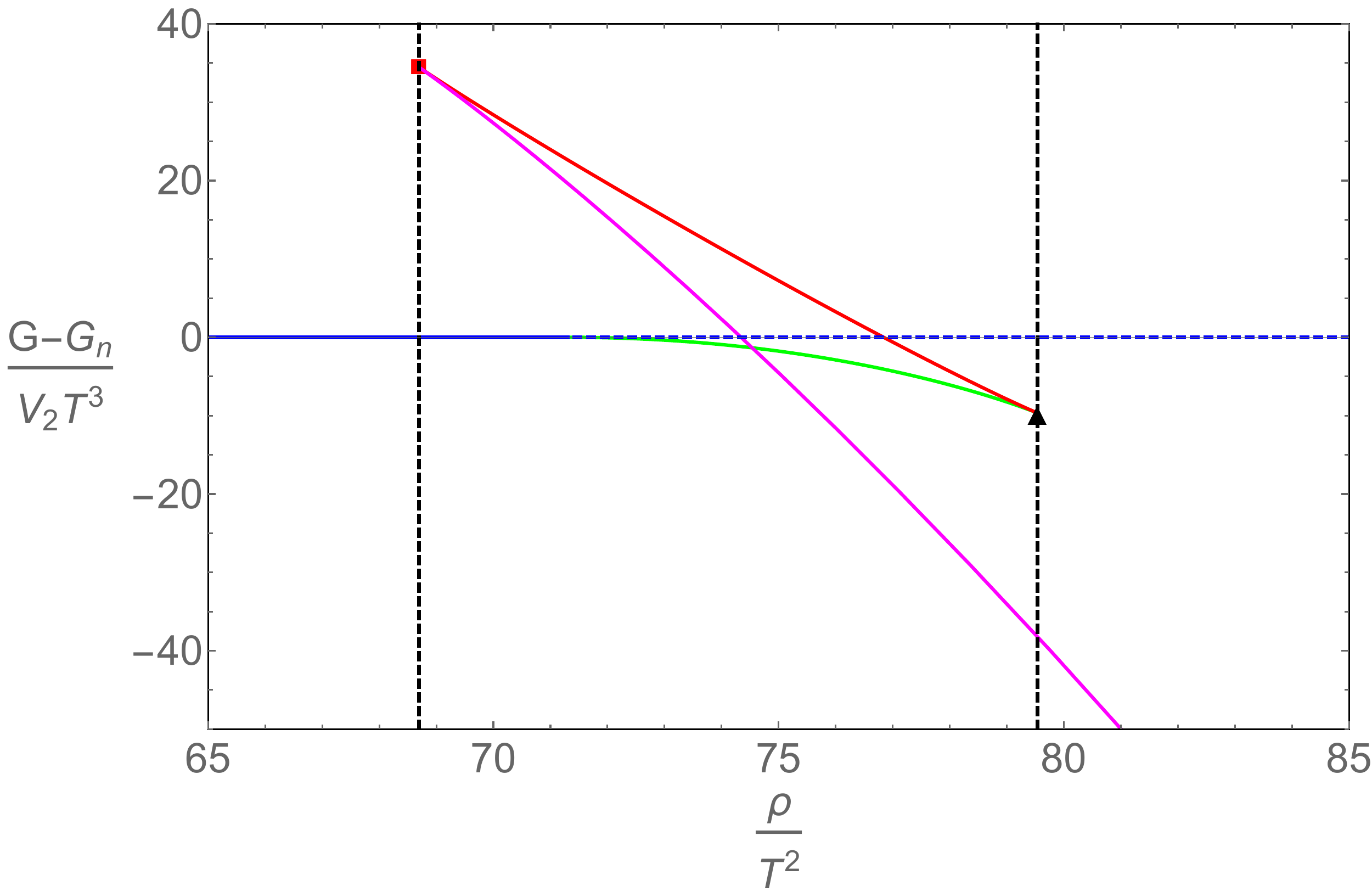}
	\caption{A typical cave of wind phase transition. \textbf{Left: } The normalized scalar condensate as a function of the dimensionless charge density. \textbf{Middle: } The imaginary part of the amplitude mode of the superfluid phase. \textbf{Right: } the {Gibbs free energy} difference between the normal phase and the superfluid phase. All figures are for $\lambda=-0.8,\tau=0.12$ and $m^2=-2$, $q=1$. The black triangle and the red square mark the two turning points. The black square denotes the position where the amplitude mode collide with another mode on the imaginary axes.}\label{condensate_free_cow}
\end{figure}

The idea is to start from the potential for the 0th order phase transition with negative $\lambda$ and zero $\tau$ and to modify it by introducing a small positive value for the $\tau$ parameter. The results for the condensate and the {Gibbs free energy} with $\lambda=-0.8$ and $\tau=0.12$ are shown in the left and right panels of Figure \ref{condensate_free_cow}. As evident from there, we have a lower branch for the superfluid solution (green curve) in which the condensate grows with the charge density. At a certain value of $\rho$, there is a turning point in which the condensate turns backwards and starts growing by decreasing the charge density (red curve). Continuing along the vertical direction, there is a second turning point where this behavior is further inverted and a top branch (magenta curve) appears in which the condensate grows monotonically towards $\rho \rightarrow \infty$. Interestingly, the lower superfluid branch has the lower {Gibbs free energy} only up to a phase transition point, after which the top branch becomes the thermodynamically favorable solution. 

\begin{figure}
	\center
	\includegraphics[width=0.6\columnwidth]{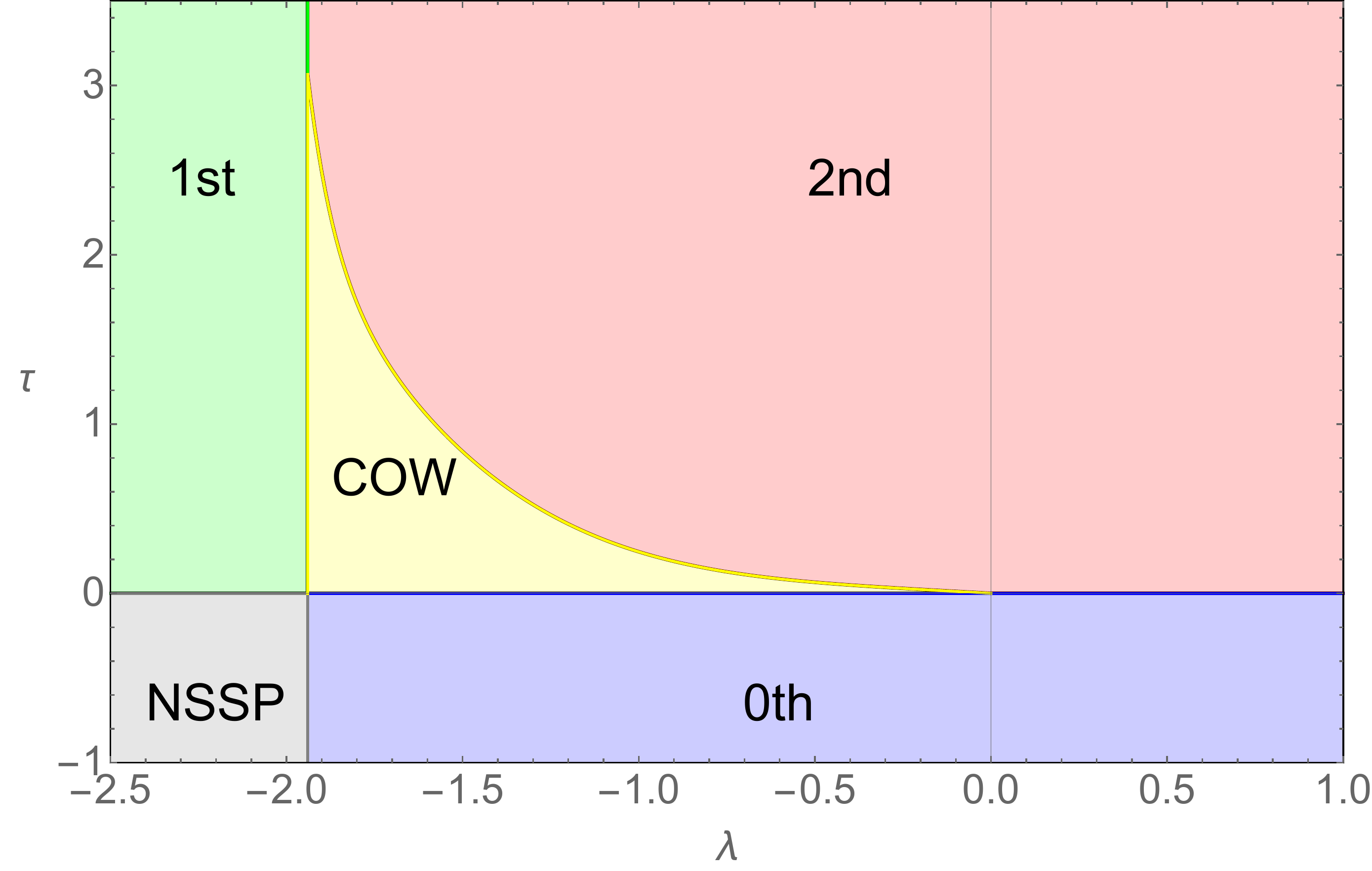}
	\caption{Phase diagram with all the possible types of phase transition as a function of the potential parameters $\lambda,\tau$. COW stands for ``cave of wind", NSSP stands for ``no stable superfluid phase''.	}\label{fig:phases}
\end{figure}

All in all, this holographic model with the nonlinear parameters $\lambda,\tau$ displays a complex phase diagram with many different types of phase transitions depending on the values of the two parameters. We provide a direct visualization of the whole shebang in Figure \ref{fig:phases}. The boundary of the 2nd order and ``cave of wind''(COW) phase transitions is obtained numerically. The boundary between the 0th and {COW} phase transitions is located at $\tau=0$ because from the landscape point of view, when $\psi$ goes to infinity, the potential goes to positive infinity for an arbitrary small positive value of $\tau$. Then, the system is rescued from the problematic 0th order phase transition, and undergoes a {``cave of wind''} phase transition. The same analysis holds for the boundary between the ``no stable superfluid phase''(NSSP) case and the 1st order phase transition. The boundary between the COW phase transition and the 1st order one, as well as the boundary between the ``no stable superfluid phase''(NSSP) case and 0th order phase transition are decided by the vertical line $\lambda=\lambda_s$, which separates the two cases where the condensate near the critical point grows leftwards and rightwards, respectively.

\subsection{The landscape analyses for the phase transitions} 
In the previous section, we have presented all sorts of phase transitions by analyzing their condensate and the behavior of the Gibbs free energy around the {phase transition} points. {Before we go to the direct study on the dynamical stability via QNMs, we speculate the \textit{landscape} pictures for these typical phase transitions from the finite information of equilibrium solutions.}

The landscape picture is popular in explaining a concrete picture for phase transitions with spontaneous symmetry breaking~\cite{doi:10.1142/0215} as well as the meta stable states in 1st order phase transitions. However, a rigorous and clear definition to the landscape method is difficult to find out in contexts. In Ref.~\cite{Li:2020ayr}, the authors give a detailed description to the landscape picture in holography. Here we just use a simplified version, defined on a subspace of homogeneous configurations. 

In order to study the dynamic stability of a physical system, we need to specify some physical conditions such as fixed total charge or chemical potential, which is related to different ensembles for equilibrium states. Under these special physical conditions, we further add small perturbations onto the solution to see whether it goes away to another state or dissipating all the fluctuations and go back to the original state. This original state is thus claimed to be dynamically unstable or stable, respectively. The landscape picture is used to present this dynamical stability in a more concrete way, with the principle that the stable equilibrium states get minimum value of thermodynamic potential in all possible configurations under the same physical conditions. In statistical physics, it is claimed that from the variation of thermodynamic potential functional, the equations of states are deduced, which means that in a state determined by the equations of states the variation of the thermodynamic potential is 0. If we plot the value of thermodynamic potentials for all possible configurations on the configuration space, we can see that all states satisfying the equations of states are stationary points in this thermodynamic potential landscape. Further more, the stable states are indeed minimum and unstable states are maximum or saddle points in this thermodynamic potential landscape, which is consistent with the information from dynamical stability.

In AdS/CFT, the boundary equilibrium homogeneous states at finite temperature is mapped to a static solution, depending only on the radial coordinate, of the bulk classical gravity system with an event horizon. It is also convenient to define a landscape picture because the classical solutions satisfy the equations of motion, which is a result of the least action principle. Therefore the bulk version of the landscape picture is defined by the value of the action of all possible configurations for the classical fields under the same boundary conditions, which are dual to the special physical conditions of the boundary thermal system. It is then expected to have a duality between the landscape picture of thermodynamic version for the boundary quantum field system and that of the least action version for the bulk classical system.

The space of all possible configurations are infinitely dimensional. In order to draw a picture of the landscape more concretely, We choose a path connecting all the stationary points on the space of configurations depending only on radial coordinate $r$, which is dual to homogeneous states in the boundary theory. In order to describe the landscape more clearly, we combine the two aspects of the duality. We use the value of thermodynamic potential functional in the boundary field theory to setup the landscape, while use some typical value of bulk fields to label the different configurations along the path.

\begin{figure}
\centering	\includegraphics[width=0.45\columnwidth]{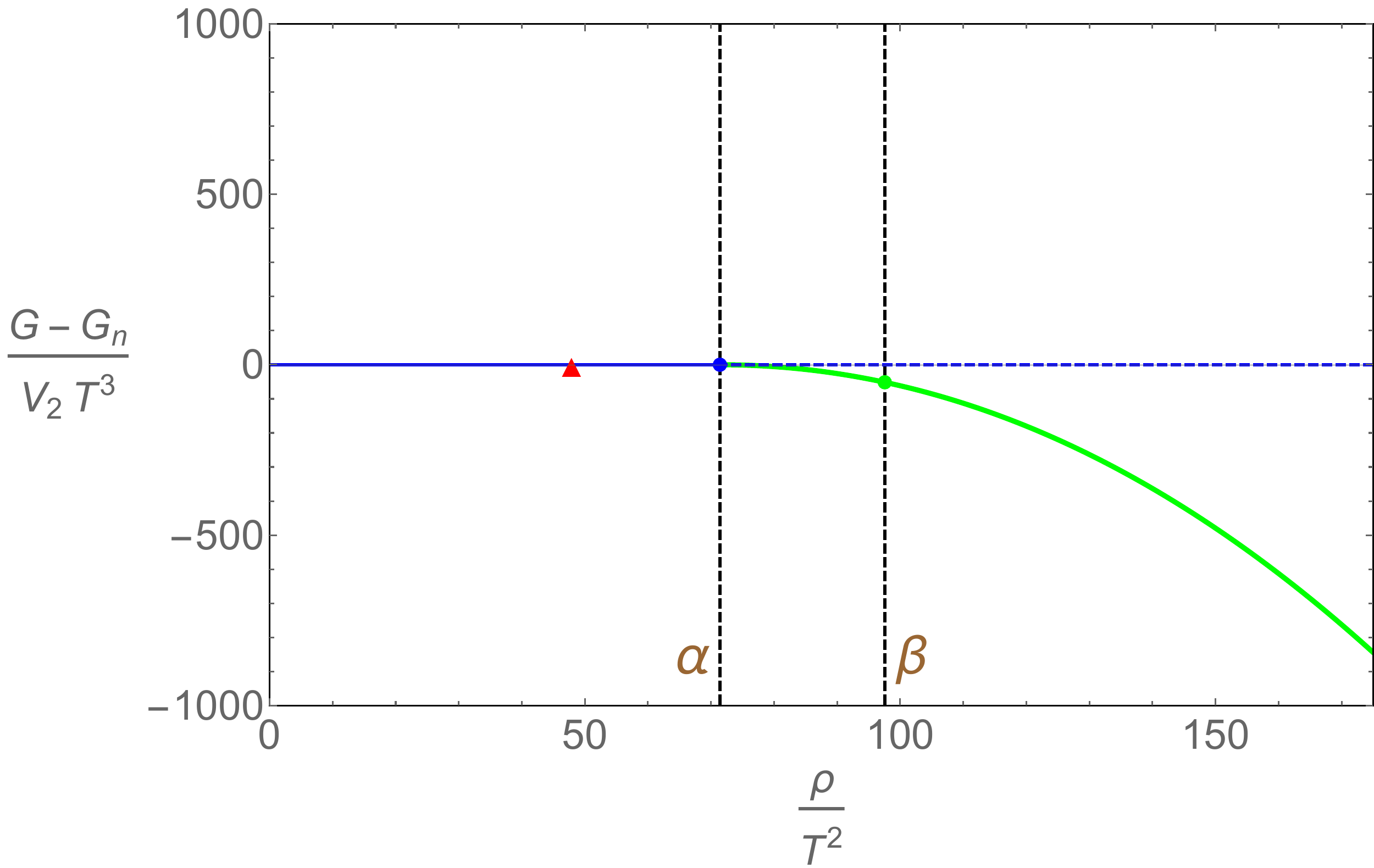}\\
		\includegraphics[width=0.45\columnwidth]{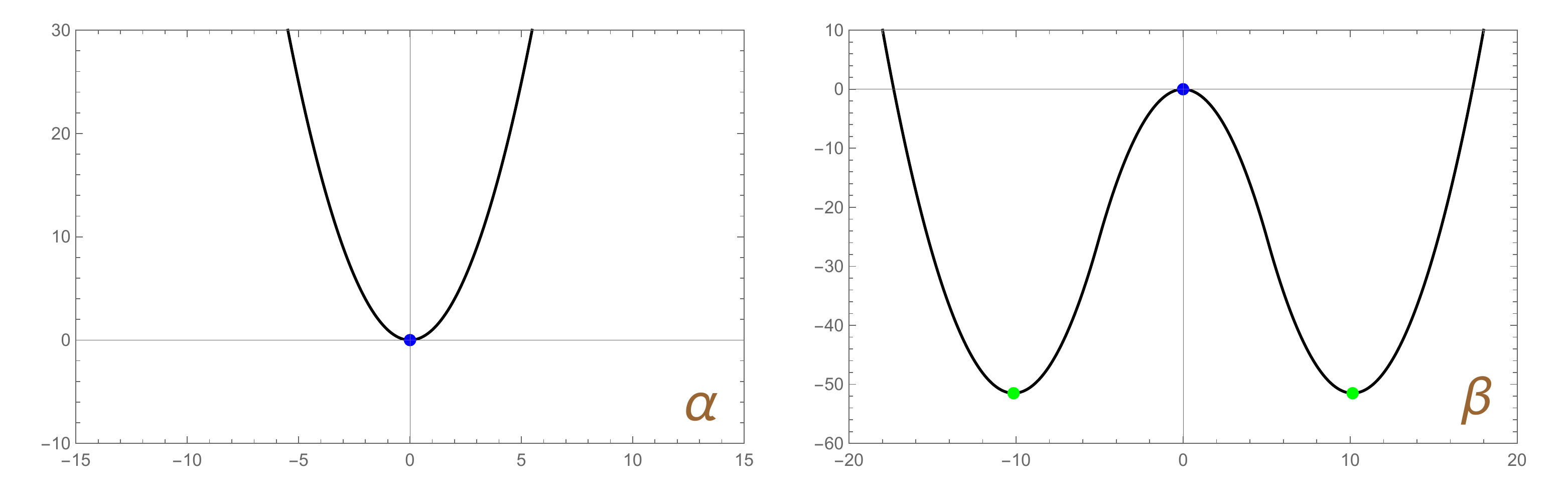}
	\caption{The landscape viewpoint for 2nd order phase transition with the typical Mexican hat potential. The two bottom panels show the potential at the critical point and in the broken phase.}\label{condensate_diagram_landscape2nd}
\end{figure}

Let us firstly take the most common 2nd order phase transition as an example. In Fig.~\ref{condensate_diagram_landscape2nd}, we draw a landscape picture with the vertical axes denoting the thermodynamic potential while the horizontal axes marking the position along the 1 dimensional path connecting the typical stationary points in the configuration space. For off shell configurations, it is hard to define thermodynamic variables or the order parameters in the boundary field theory, however, in the classical bulk system, the boundary values of the bulk fields are still clear. Therefore a convenient choice of the horizontal axes is the boundary coefficient $\psi^{(2)}$ of the scalar field $\psi$. Because we choose to work in canonical ensemble and show Gibbs free energy as function of charge density $\rho$ for different on-shell solutions, we expect the resulting landscape pictures give the correct dynamic stability with the physical condition of fixed total charge. We should notice that when we work in the subspace of only homogeneous states, fixed total charge is equivalent to fixed charge density $\rho$.

From a fixed value of $\rho$ in the plot of Gibbs free energy, we count how many different kinds of solutions we have. When $\rho<\rho_c$, the system is in the normal phase, which is also the only solution in the plot of 2nd order phase transition. Therefore we speculate that there is only a single stationary point which is located at the origin. Above the critical point with $\rho>\rho_c$, there are two solutions in the plot of Gibbs free energy, therefore besides the normal solution with $\psi=0$, there is also a solution with finite scalar field $\psi$. The normal solution get a higher value of Gibbs free energy and is a maximum in the landscape, while the solution with nonzero $\psi$ is a minimum. Because the U(1) symmetry, this finite $\psi$ solution is degenerate with other cousins with a nonzero phase angle. For simplicity, we only show the two symmetric minima with $0$ and $\psi$ phase angles on the 1 dimensional configuration space. We can rotate the curve around the vertical axes and get the typical Mexican hat potential related to spontaneous symmetry breaking. The U(1) symmetry and the degeneracy of finite $\psi$ solutions also exist in the following landscapes for other types of phase transitions. It is necessary to be clarified that we haven't really calculate the off shell configurations, but speculate the qualitative landscape from only the information of thermodynamic potential of on shell configurations. Therefore the curves presenting the landscape is accurate only at stationary points.

\begin{figure}
\centering
\includegraphics[width=0.5\columnwidth]{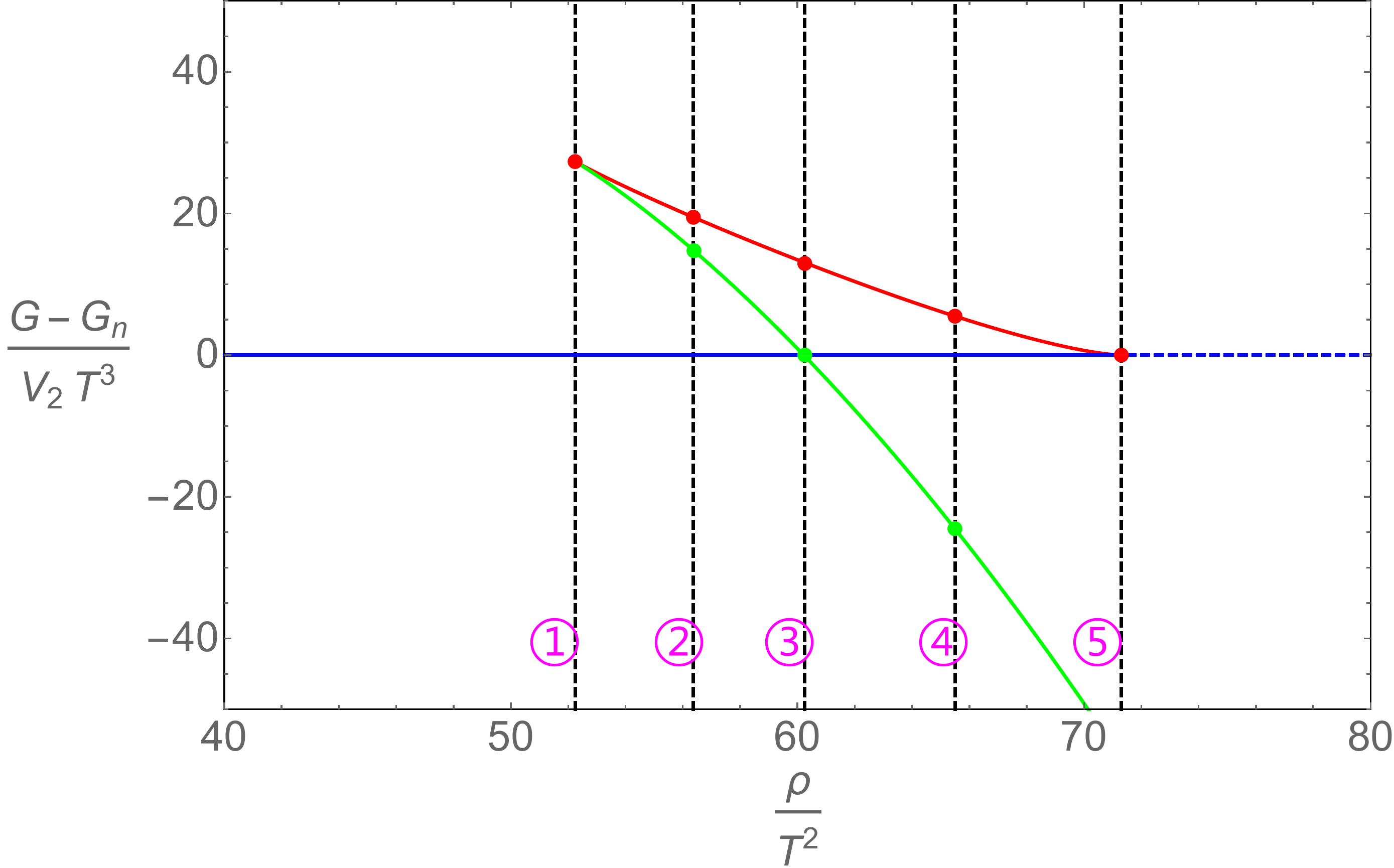}\\
	\includegraphics[width=0.5\columnwidth]{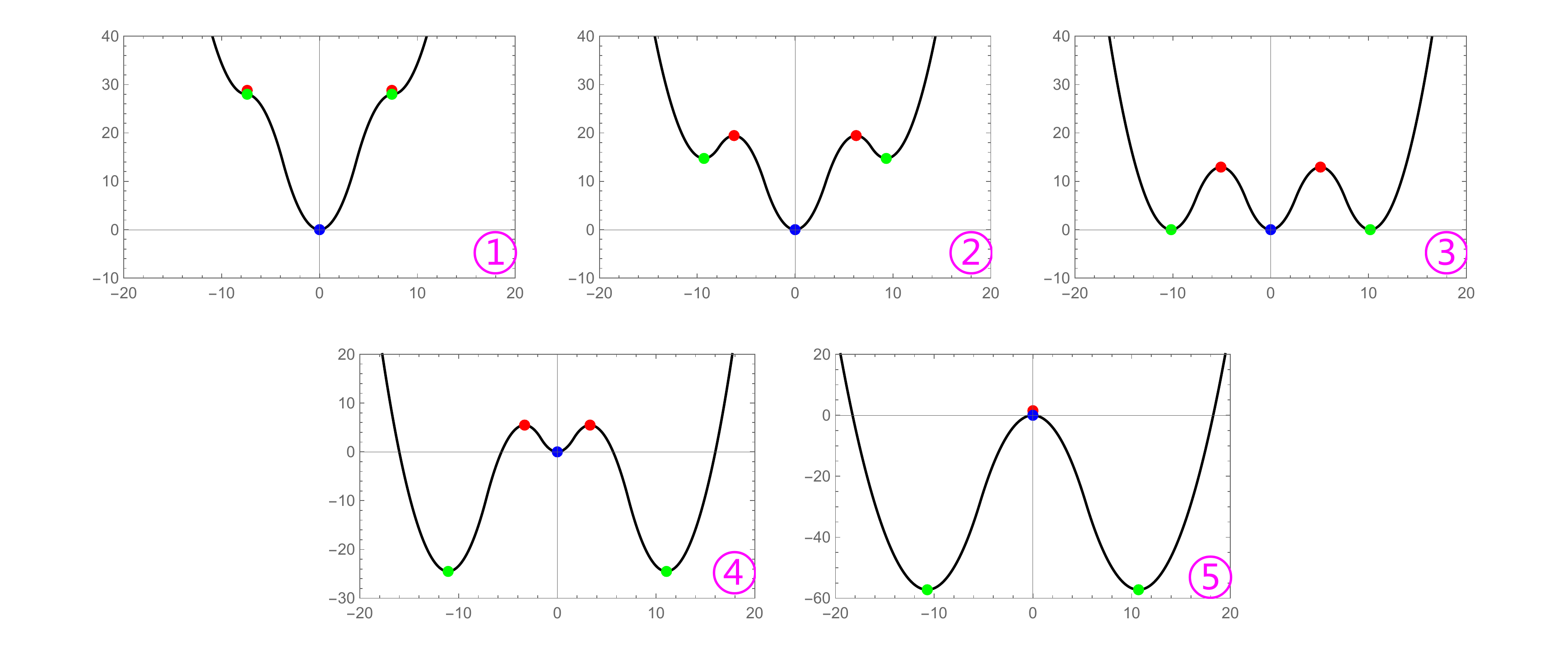}
	\caption{The potential landscape the 1st order phase transitions. The shapes of the potential indicated with \ding{172}$-$\ding{176} relate to the corresponding points in the top panel. Same colors are also used.
	}\label{condensate_diagram_landscape1st}
\end{figure}
Let us now move to the case of a 1st order phase transition and show the corresponding landscape viewpoint in Fig.\ref{condensate_diagram_landscape1st}. For simplicity, we choose five benchmark points and we show the {Gibbs free energy} landscape for each of them (bottom panels in Fig.~\ref{condensate_diagram_landscape1st}). We analyze from low charge density to high charge density, which corresponds to moving from point \ding{172} to point \ding{176}. Below point \ding{172}, there is only one solution at a fixed value of $\rho$, and there is only one minimum at origin in the landscape. Right at point \ding{172}, besides the global minimum at the origin, an additional solution related to the tip point on the swallow tail arise. This is a secondary saddle point indicated in red and green, because it is split to two solutions when $\rho$ is further increased. Again because of the U(1) symmetry, we have the other secondary saddle point on the other side of the 1-dimensional configuration space.

By moving further, in point \ding{173}, the two secondary saddles points split into a pair of maxima (red) and a pair of local minima (green). Still the minimum at the origin has a lower Gibbs free energy and is the thermodynamically favorable equilibrium solution. For a special value of $\rho=\rho_T$, which is indicated in point \ding{174}, the Gibbs free energy of the local minima marked green becomes equal to that of the minimum at the origin. This is the location of the first order phase transition in common sense. Above this phase transition point, see \ding{175}, the minima shown in green become the thermodynamically favorable solution. At the phase transition point, the order parameter jumps from zero (blue point) to a finite value given by the location of the green point. As expected, the phase transition is discontinuous. At point \ding{175}, we still observe two local maxima (indicated with red) that will eventually collapse into the origin by increasing further the charge density (see point \ding{176}).

\begin{figure}
\centering
\includegraphics[width=0.5\columnwidth]{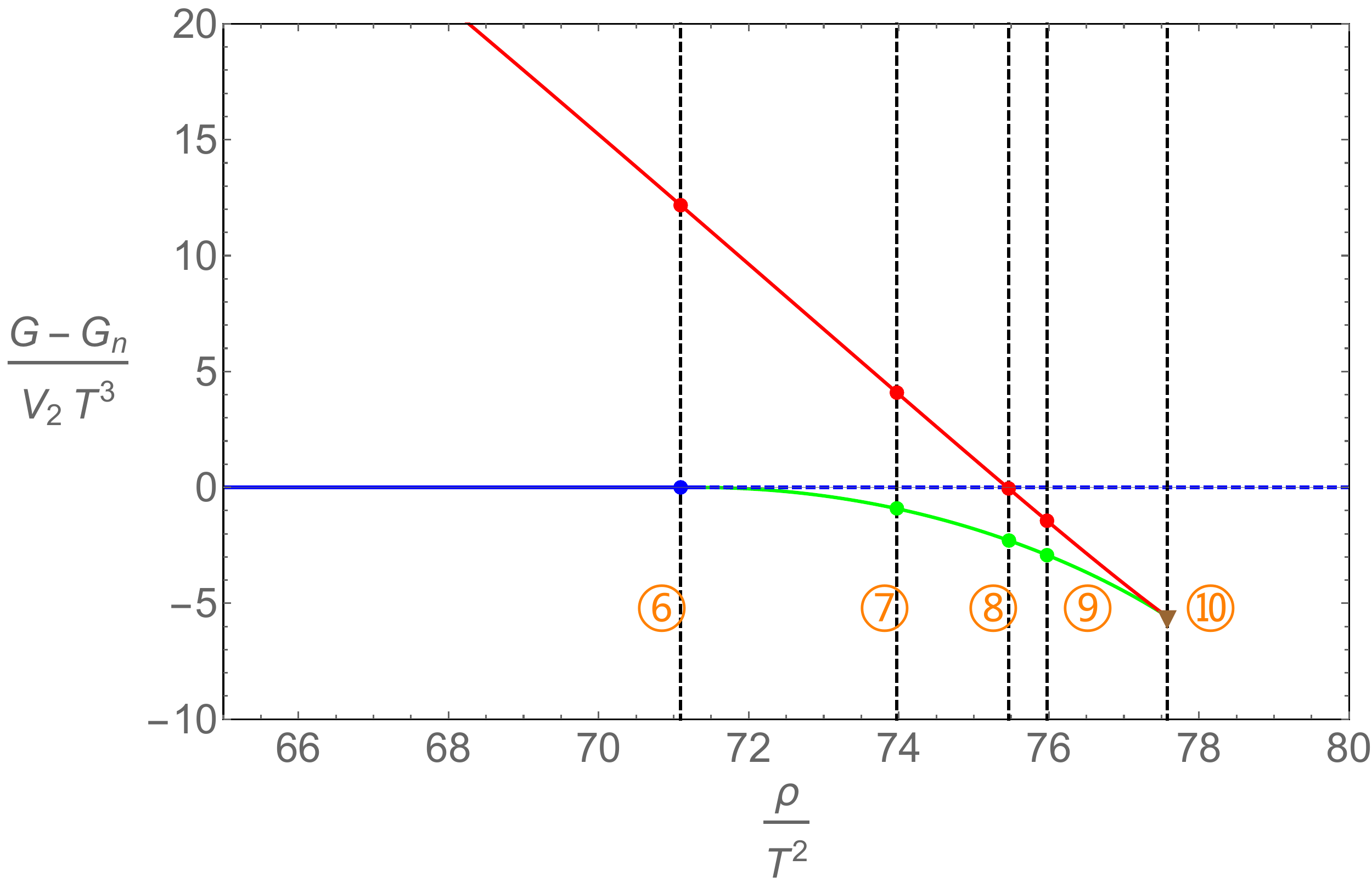}\\
\includegraphics[width=0.5\columnwidth]{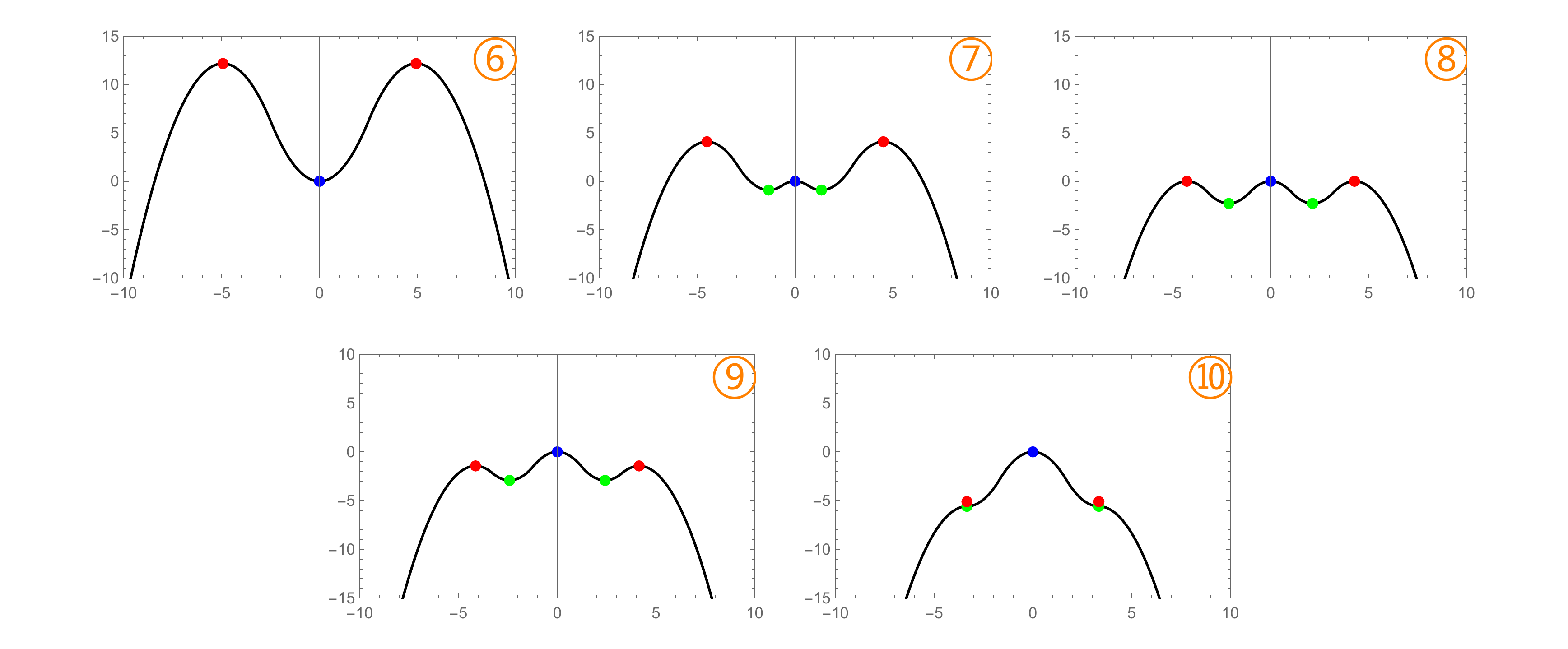}
	\caption{The potential landscape the 0th order phase transitions. The shapes of the potential indicated with \ding{177}$-$\ding{181} relate to the corresponding points in the top panel. Same colors are also used.
	}\label{condensate_diagram_landscape0th}
\end{figure}

Let us now move to the 0th order phase transition whose landscape is pictured in Figure \ref{condensate_diagram_landscape0th}. For clarity, we have isolated five benchmark points, \ding{177}-\ding{181}, to discuss the whole dynamics.  At small charge density, below point \ding{177}, there always exist two global maxima in the potential (red) together with the meta-stable minimum located at the origin (blue). At the critical point, located at point \ding{177}, the potential displays two new minima indicated in green, while the initial minimum located at the origin becomes a maximum. The two new minima and the central maximum with lower {thermodynamic potential} are well visible in point \ding{178}. By moving further, in point \ding{179}, all the global maxima (red and blue) have the same zero {thermodynamic potential} and the minima are still located at the green points. Continuing in this direction, in point \ding{180}, the {thermodynamic potential} of the pair of local maxima (red) becomes smaller than the that of the global maximum at the origin (blue). Finally, at the turning point corresponding to point \ding{181}, the lowest maxima (red) merge with the minima (green) and form unstable saddled points. Above the value of the charge density of the turning point, the potential has no minima but only one global maximum at the origin.

Notice that the potential is not bounded from below and it diverges in the limit of $|\psi|\rightarrow \infty$ for all the values of charge density $\rho$, therefore the system is always unstable from the global thermodynamic point of view. Even if we assume all the meta-stable states to be long lived, there is no minimum with $\rho>\rho_t$. Therefore the so called 0th order phase transition with the system jumping from a solution with lower value of thermodynamic potential to another one with higher value of thermodynamic potential is impossible to occur, from the land scape point of view. It is also easy to check by calculating the QNMs of these different solutions to confirm the local stability. This is a main motivation of this work at the beginning.

Analyzing further the influence of the quartic term with coefficient $\lambda$ is helpful to understand how the landscape change to the style of ``0th order phase transition''. A positive $\lambda$ corresponds to a repulsive self-interaction, while a negative one to an attractive term in the potential. In presence of attractive self-interactions, the potential diverges to negative infinity for infinitely large value of $\psi$. Then the system becomes unstable from the global thermodynamic point of view. Furthermore, when there is a local minimum in the landscape, there must be a saddle point somewhere between this local minimum and the point at infinity. This saddle point is nothing but the unstable branch of superfluid solutions with higher value of condensate, which is in red color in the left panel of Figure.~\ref{condensate_free_0th}.

\begin{figure}
	\center
	\includegraphics[width=0.6\columnwidth]{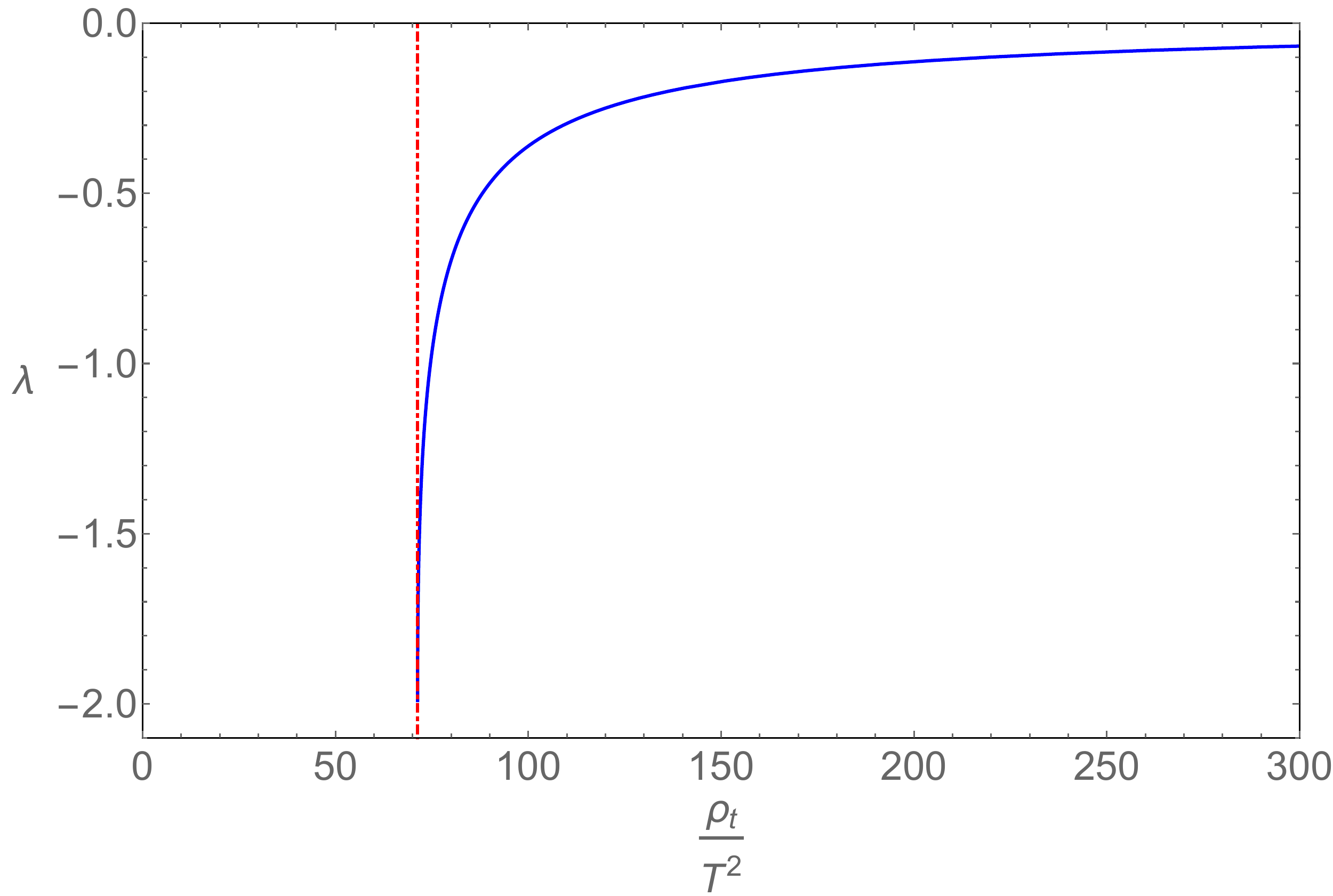}
	\caption{The dependence of the turning point $\rho_t$ as a function of the parameter $\lambda$ in the 0th order phase transition case. For an example, see the left panel of Figure \ref{condensate_free_0th}.
	}\label{Ls-rho-PD}
\end{figure}

Before moving to the next and last case, we want to discuss in more detail the role of the parameter $\lambda$. In Figure~\ref{Ls-rho-PD}, we draw a $\lambda-\rho$ phase diagram in which the position of the turning point is indicated. In this plot, we can see two curves. The vertical dashed line marks the critical point at which the superfluid phase transition takes place. As studied in Ref.~\cite{Zhang:2021vwp} and explained above, the value of $\lambda$ does not modify the location of the critical point. The blue line shows the dependence of the turning point $\rho_t$ of the zeroth order phase transition as a function of the parameter $\lambda$. The position of the turning point reaches the critical point at $\lambda=-1.949$. Below that point, the condensate starts growing leftwards, as already observed in Ref.~\cite{Herzog:2010vz} with 4+1 dimensional bulk using analytical perturbative techniques.

In Ref.~\cite{Herzog:2010vz}, it was further claimed that when the superfluid solution moves in the opposite direction near the critical point, the system meets with either ``\textit{a first order phase transition or runaway pathologies}''. Our study shows that, at least in the four-dimensional bulk with the $\lambda$ term, the correct answer is: ``runaway pathologies". However, the answer can also be changed to ``a first order phase transition'' with an additional positive value of $\tau$. We expect these result to be universal and also applicable to more general models with $\psi^4$ and $\psi^6$ terms, in light of the following analysis.

Because of the ``runaway pathologies'', the system with negative value of $\lambda$ and no higher order term is generally unstable. However, with a positive higher order term which dominates the area in the landscape with very large value of $\psi$, the system will become globally stable again. In this work, we add a sixth order term with coefficient $\tau$ as introduced in the action. This higher order term would not affect the qualitative behavior near the critical point $\rho=\rho_c$. Therefore, with an additional positive $\tau$, we get a system with 1st order superfluid phase transition with $\lambda<\lambda_s$ and a system with ``cave of wind'' (or 2nd order) phase transition with $\lambda>\lambda_s$. The landscape for the 1st order superfluid phase transition has just been introduced, we now conclude this section by analyzing the landscape of the cave of wind phase transition.

\begin{figure}
\centering
\includegraphics[width=0.5\columnwidth]{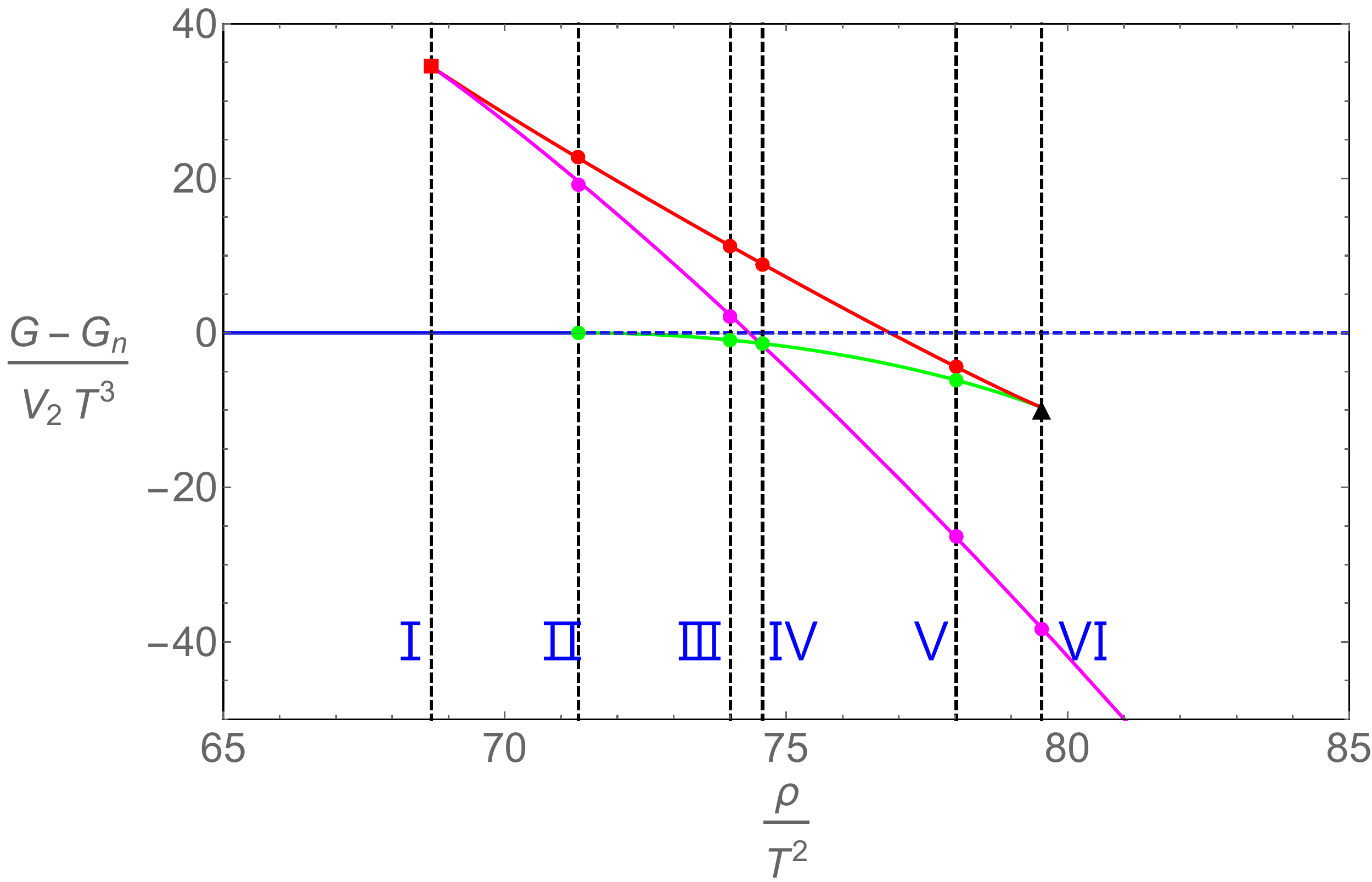}\\
\includegraphics[width=0.5\columnwidth]{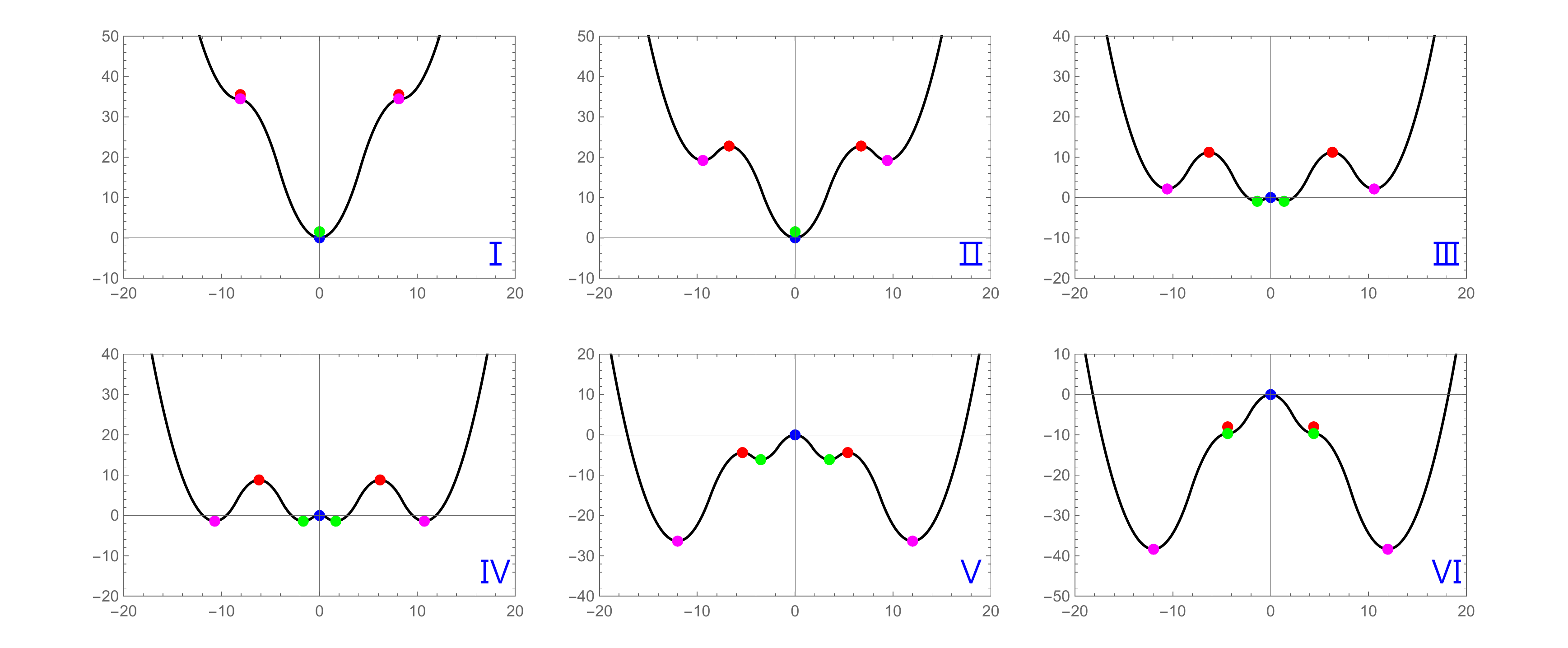}
	\caption{The potential landscape the cave of wind phase transition. The shapes of the potential indicated with (I)-(VI) relate to the corresponding points in the top panel. Same colors are also used.
	}\label{condensate_diagram_landscapeCOW}
\end{figure}

We show the results in Fig.~\ref{condensate_diagram_landscapeCOW}. Starting from low charge density and a single global minimum at the origin (blue), the system first develops a double copy of saddle points (red and magenta) which are visible in point (I). After that, this pair of saddle points are split into a pair of local minima(magenta) with higher {Gibbs free energy} than the global one (blue), and a pair of saddle points(red). At the critical point, (II), the system exhibits an additional pair of minima (green) whose {Gibbs free energy} is now lower than all other stationary points, see for example (III). At that moment, the stationary point situated at $\psi=0$ becomes a maximum. By continuing in this direction, in point (IV), the pair of local minima at larger value of the condensate (magenta) has the same {Gibbs free energy} as the one with smaller condensate. By increasing even further, in point (V), these two external minima (magenta) become global minima and thermodynamically favorable solutions, indicating a 1st order phase transition between the two superfluid solutions at point(IV). Finally, for even larger charge density, in point (VI),  the minima with higher {Gibbs free energy} coalesce with the saddle point and disappear together for higher values of the charge density. From the landscape point of view, it is clear that when $\tau$ is large enough, the red saddle points and the magenta minima will disappear because the positive $\tau$ term dominates in finite condensate solutions, this is reason that the 2nd order phase transition also dominates the upper region with $\lambda_s<\lambda<0$ in the phase diagram in Fig.~\ref{fig:phases}.

\section{Dynamical stability from quasi normal modes}\label{sec3}
\subsection{Homogeneous perturbations \& $k=0$ quasi normal modes}
The linear dynamical stability of a system can be investigated by looking at the spectrum of low-energy excitations.\footnote{More precisely, an analysis of the low-energy excitations reveals only instabilities which might happen in that regime and could miss other more microscopic instabilities encoded in the dynamics of the far-away non-hydrodynamic modes active at large frequency and wave-vector.} In presence of dissipation and finite temperature, these excitations have in general a complex dispersion relation and they are referred to as ``quasi normal modes". More in general, each of them obeys a relation between the complex frequency $\omega$ and the real wave-vector $k$ which is given by:
\begin{equation}
    \omega(k)= \mathrm{Re}(\omega)+i\, \mathrm{Im}(\omega)\,.
\end{equation}
QNMs control the relaxation towards equilibrium and the late time dynamics.
Dynamical stability is just the statement that the imaginary parts of all the quasi normal modes are not positive, $\mathrm{Im}(\omega)\le 0$. If that were not the case, then the system would suffer of an exponential unstable growth $\propto \exp(\mathrm{Im}(\omega) t)$. If this instability appears at zero wave-vector, $k=0$, we call it a homogeneous instability. If on the contrary, the imaginary part becomes positive at a finite critical wave-vector, then we have an inhomogeneous instability.

From the gravity side, quasi normal modes can be directly computed from the equations of motion for the linearized bulk perturbations \cite{PhysRevD.72.086009}. In this section, we calculate the quasi normal modes for the different types of phase transitions discussed in the previous sections using the same holographic model. For simplicity, we focus on the lowest lying quasi normal modes. Also, in this subsection, we will discuss only the homogeneous case. The dynamics at finite wave-vector, $k\neq 0$, will be analyzed later.

\begin{figure}
	\center
	\includegraphics[width=0.6\columnwidth]{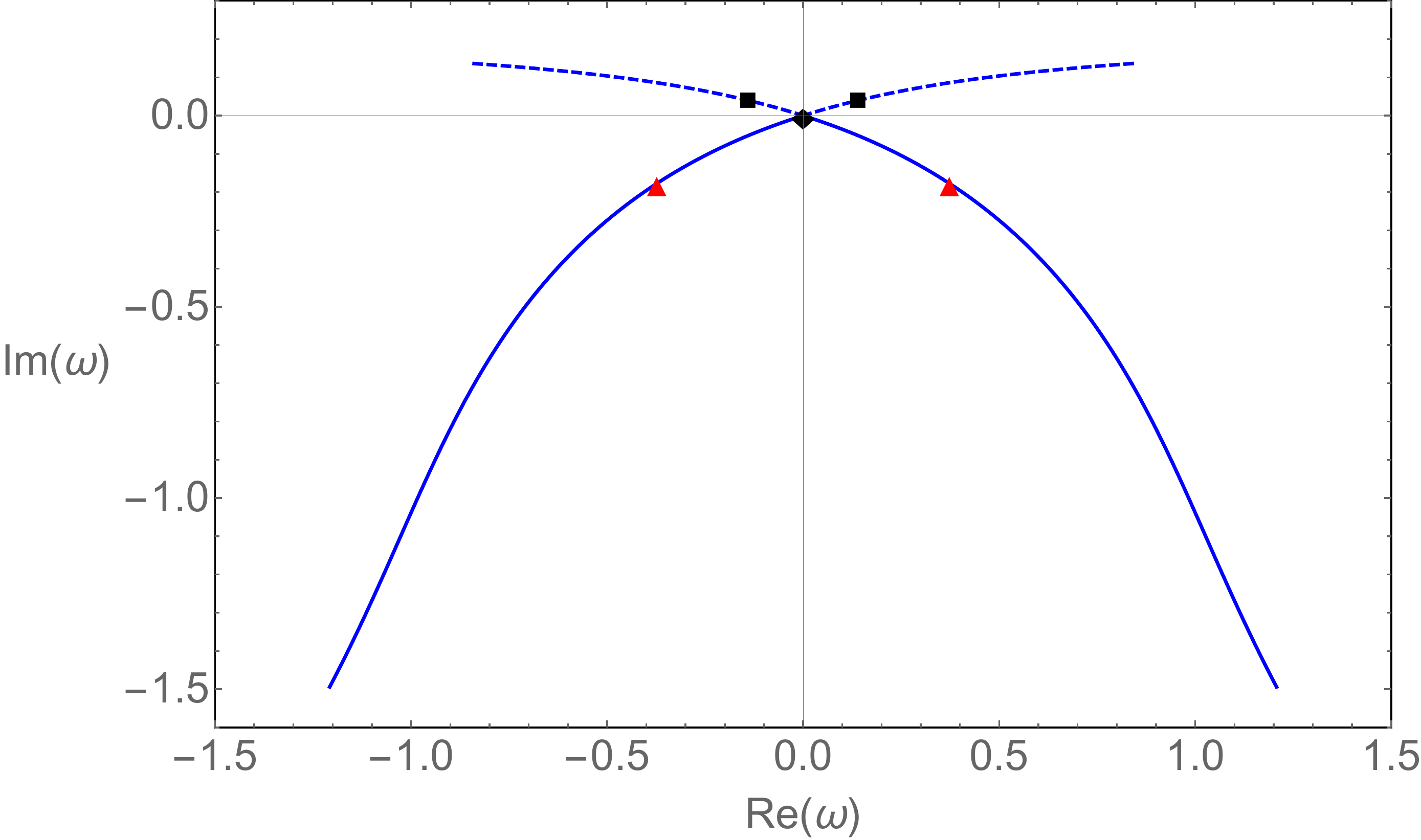}
	\caption{The behavior of the two lowest lying scalar quasi normal modes in the normal solution moving across the 2nd order phase transition. The rhombus indicates the position of critical point.}\label{QNMsN}
\end{figure}

As a warm up, let us start by reviewing the most familiar case of the 2nd order phase transition. The original results can be found in \cite{Amado:2009ts}. We represent the results of the most important scalar modes for the normal solution in Fig.~\ref{QNMsN}, with the black rhombus indicating the position of the critical point. In the normal phase, we have two critical scalar modes with both real and imaginary frequency together with a mode from the gauge sector which lies at the origin for $k=0$. The latter is decoupled from the scalar modes and it corresponds just to standard charge diffusion. By approaching the critical point, the two lowest lying stable modes move towards the origin of the complex frequency plane. Exactly at the critical point, all the three modes sit at the origin. If $\rho$ is continuously increased above the critical value, $\rho>\rho_c$, the two lowest lying modes move into the upper half of the complex plane after crossing each other. This is a clear signal of instability suggesting that the system will be deformed into another (possibly stable) phase.

The broken superfluid phase is stable above the critical point, for $\rho>\rho_c$. There, one of the scalar quasi normal modes remains at the origin while the other moves down the imaginary axes. The mode remaining at the origin is the Goldstone mode and corresponds to the phase of the order parameter. The second mode is the amplitude mode, which is not a hydrodynamic mode. In the middle plot of Fig.~\ref{condensate_free_2nd}, we show the dependence of the imaginary part of the amplitude mode as a function of the condensate. We take the condensate as the vertical axes for better comparing with the condensate plot in the cases with turning points. Notice that the behavior of this mode close to the critical point can be derived analytically as recently shown in \cite{Donos:2022xfd}. Moreover, the imaginary part of this mode goes to zero at the critical point. This produces interesting physical consequences such as the fact that the relaxational dynamics exactly at the critical point is governed by the full nonlinear dynamics resulting in a power law rather than exponential decay \cite{Flory:2022uzp,Cao:2022mep}.

We now turn on the $\lambda$ and $\tau$ parameters in order to study the quasi normal modes for 1st and 0th order phase transitions. It is worth noticing that these new terms are all nonlinear in the scalar field $\psi$. As a consequence, they will not modify the {QNMs for the normal solution, which is determined by linear dynamics on the background solution with $\psi=0$.}

\begin{figure}
	\center
	\includegraphics[width=0.3\columnwidth]{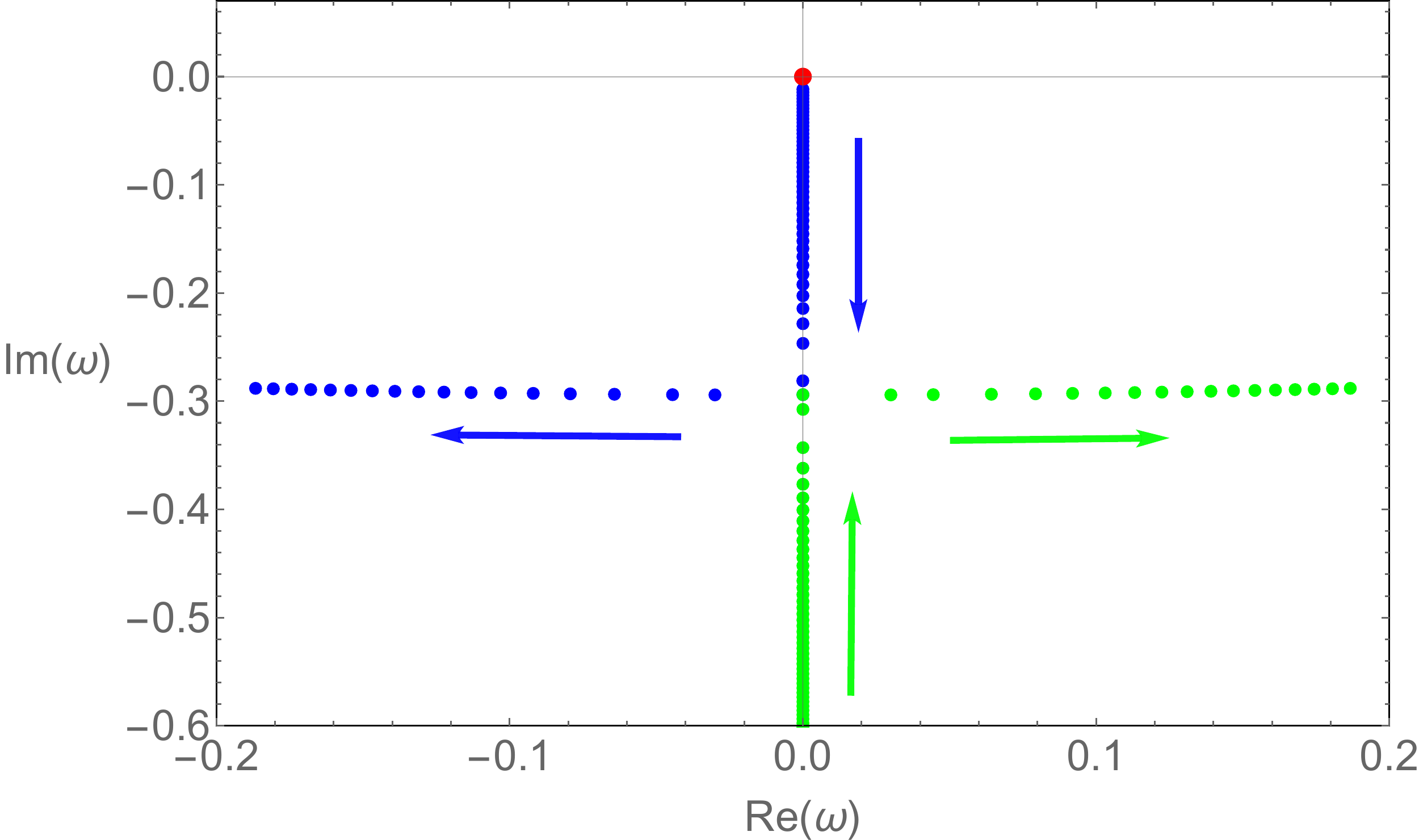}
    \includegraphics[width=0.3\columnwidth]{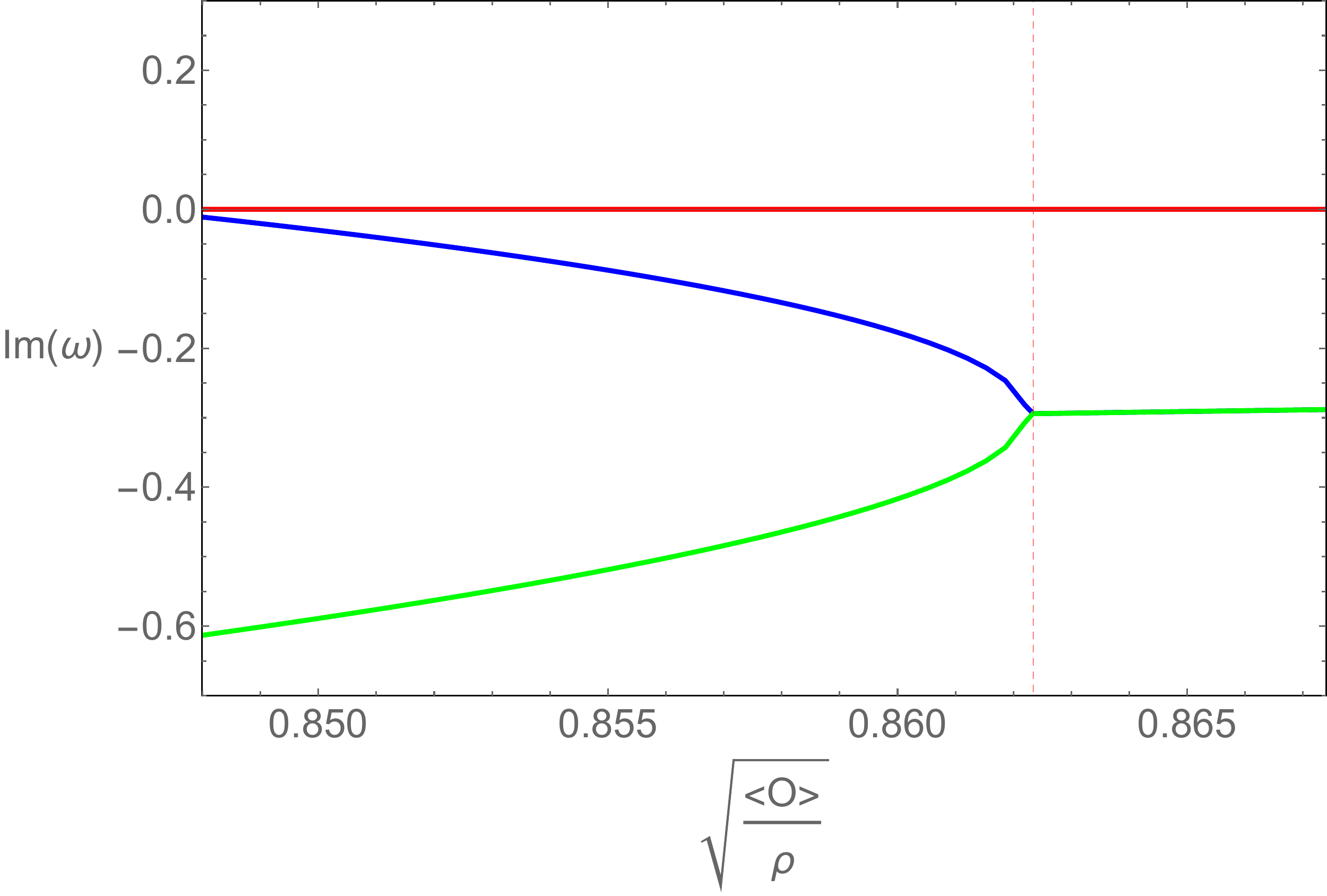}
    \includegraphics[width=0.3\columnwidth]{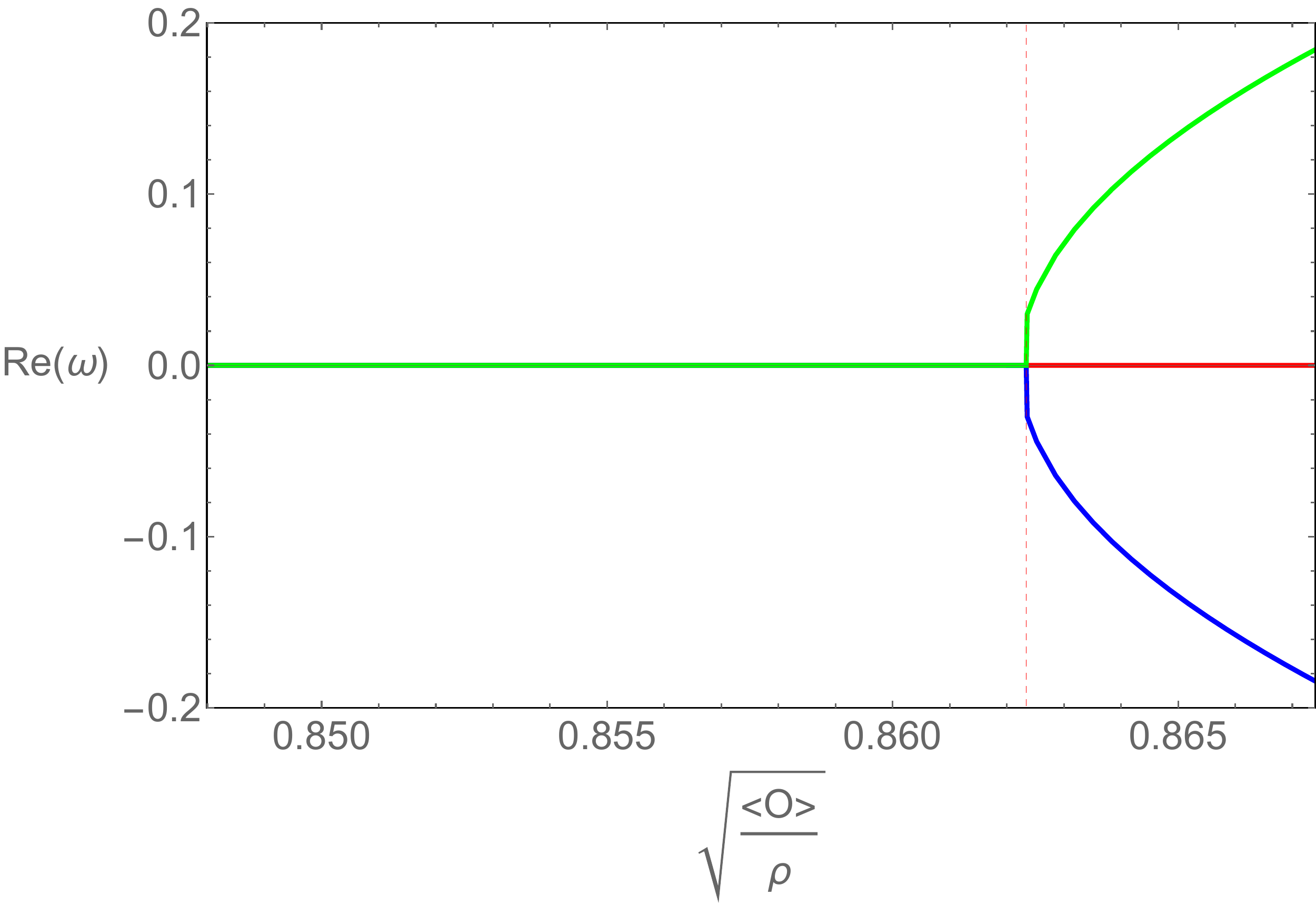}
	\caption{The dynamics of the amplitude mode in the 1st order phase transition in the broken phase. The blue line is the trajectory of the amplitude mode shown in the middle plot of Figure~\ref{condensate_free_1st}. The green line is an additional non-hydrodynamic mode which originally sits on the negative axis. The red line is the Goldstone mode. The location of the collision is indicated with a dashed line in the \textbf{middle and right} panels. The \textbf{left} panel shows the dynamics in the complex frequency plane, making clear the collision between the blue and green modes.
	}\label{condchange_omega_rho}
\end{figure}

We start with the case of the 1st order phase transition with $\lambda=-2$ and $\tau=0.8$. Our results indicate that one of the two critical scalar modes, the Goldstone mode, still stays at the origin of the complex plane. The other amplitude mode has zero real frequency and moves along the imaginary part as depicted in the middle plot of Fig.~\ref{condensate_free_1st}. Differently from the case of the 2nd order phase transition, this time, the imaginary part of this mode can become positive. In particular, on the lowest unstable branch shown in red color in Figure~\ref{condensate_free_1st}, the amplitude mode first moves on the positive side of the imaginary axes. This confirms the presence of a dynamical instability in perfect agreement with the landscape picture of the previous sections. At a certain point, in the middle of the unstable branch, the amplitude modes moves backwards towards the origin of the complex plane. Finally, at the turning point, the mode passes through the origin and after that its imaginary part keeps growing on the negative side of the imaginary axes. This, once again, proves the dynamical stability of the upper branch shown in green color in the left panel of Figure \ref{condensate_free_1st}. It is important to notice that the amplitude mode will not move indefinitely towards negative infinity. On the contrary, it undergoes a collision with another mode which was originally on the negative axis after which two symmetrical modes appear. This behavior is shown in Fig.~\ref{condchange_omega_rho}. It is interesting to observe that this collision does not seem to happen in the case of the 2nd order phase transition. Finally, we conclude that the lessons about dynamical stability which we have learned from the quasi normal modes are consistent with our landscape analysis in the previous section.

Let us now focus on the quasi normal modes for the 0th order phase transition. First, the Goldstone mode is still sitting at the origin of the complex frequency plane in the broken phase. Then, we turn to the study of the amplitude mode, which in this case is still a purely imaginary mode. We plot the imaginary part of the amplitude mode as a function of the condensate in the middle panel of Fig.~\ref{condensate_free_0th}. In the lower branch of the broken phase (green curve in the left panel of Fig.\ref{condensate_free_0th}) its imaginary part first goes down on the negative imaginary axes by increasing the condensate. After reaching a minimum value, the amplitude mode inverts its path and moves towards the origin of the complex plane along the imaginary axes. At the turning point, the imaginary part of the amplitude mode is zero and grows monotonically above zero on the upper branch indicated in red color in Fig.\ref{condensate_free_0th}. This confirms that the upper branch in Fig.\ref{condensate_free_0th} is also dynamically unstable.

Now we collect the results of quasi normal modes for both the superfluid solution and the normal solution. The result of quasi normal modes in the superfluid phase shows that the lower branch is stable while the upper branch is not. However, when $\rho$ reaches its turning point value $\rho_t$ and go on increasing, the system does not jump from the superfluid phase back to the normal phase because the QNMs show that this only solution at $\rho>\rho_t$ is unstable. This is consistent with the landscape which do not change discontinuously at the point $\rho_t$. Because the QNMs of the normal phase with different values of $\lambda$ and $\tau$ are all the same, the results of the scalar modes in Figure.~\ref{QNMsN} indicate instability for normal phase at the whole region $\rho>\rho_t$ for all the different kinds of phase transitions we presented in this model.

In summary, we conclude that a 0th order phase transition cannot occur. Are we always facing a global instability of the whole system, or can we rescue the unstable system with 0th order phase transition somehow? The answer is the latter. The idea is to add a positive $\tau$ term to modify the landscape in a way that it becomes bounded from below. In that way, stable solutions with lowest value of the thermodynamic potential always exist. As presented in the previous section, with a small positive $\tau$ term, we get a 1st order superfluid phase transition with $\lambda<\lambda_s$ and a ``cave of wind" phase transition with $\lambda_s<\lambda<0$.

We also studied the QNMs of the ``cave of wind" phase transition and show the imaginary part of the amplitude mode in the middle plot of Fig.~\ref{condensate_free_cow}. The main difference with the results of 0th order phase transition is that the imaginary part of the amplitude mode will not grow monotonically towards large values of the condensate but at a certain point will turn around and start decreasing again. At the turning point, indicated with a red square in the middle plot of Fig.~\ref{condensate_free_cow}, the imaginary part of the amplitude mode has zero value. After that, on the upper branch (magenta curve), the amplitude mode has a negative imaginary part. This upper branch is therefore dynamically stable. There is also a collision of the amplitude mode with another purely imaginary mode when the amplitude mode goes on moving downwards and a pair of symmetrical modes appear, which is similar as the case of 1st order superfluid phase transition.

In conclusion, the analysis of the quasi normal modes at zero wave-vector, $k=0$, is in perfect agreement with the thermodynamic potential landscape for homogeneous solutions at fixed total charge, the so-called ``canonical ensemble". As we will shortly see, this will not be the case anymore when considering perturbations at finite wave-vector.

\subsection{Dynamical stability at finite wave-vector}
In the previous section, we have analyzed in detail the behavior of the quasi normal modes in the normal and broken phase at zero wave-vector, $k=0$.
That analysis was fundamental to ascertain the dynamical stability of the various solutions but it is limited to the case of homogeneous perturbations. The dynamics of superfluids at finite wave-vector is rich and determined by its hydrodynamic description \cite{Herzog:2011ec}. In the probe limit, in the broken phase, there are two types of low-energy excitations. First, one has the so-called \textit{second sound} which is composed of the superfluid Goldstone, the phase of the order parameter, and of charge fluctuations. Second, there is a non-hydrodynamic mode which has a pseudo-diffusive dispersion relation and corresponds to the amplitude mode, the fluctuations of the amplitude of the order parameter. In the probe limit, the dynamics of these modes in the holographic superfluid model with 2nd order phase transition was shown in \cite{Amado:2009ts}. The behavior of the low-energy modes have been studied with full backreaction in \cite{Arean:2021tks} and in presence of a soft explicit breaking of the U(1) symmetry in \cite{Ammon:2021pyz}. The analysis involving also the 1st order phase transition type was done in \cite{Janik:2016btb}. Perturbative computations have also appeared in the literature \cite{Herzog:2010vz,Donos:2021pkk,Donos:2022xfd,Donos:2022www}. Finally, the dynamics of the order parameter has been investigated numerically in \cite{Plantz:2015pem,She:2011cm} and analytically in \cite{Donos:2022xfd,Donos:2022qao}.
	
In this section, we aim to study the behavior of the lowest quasi normal modes across the different types of phase transition at finite wave-vector, $k\neq 0$. In particular, we are interested in the dispersion relation of the second sound mode. In the superfluid phase, second sound disperses as:
\begin{equation}\label{ss}
    \omega=\pm v_2 k-i\frac{\Gamma_2}{2}k^2+\dots
\end{equation}
with $v_2$ its sound speed and $\Gamma_2$ its attenuation constant (see for example \cite{Arean:2021tks}). Dynamical stability requires that:
\begin{equation}
    v_2^2\,>\,0\,,\qquad \Gamma_2\,>0\,.
\end{equation}
We do expect that the dynamical stability of superfluid solutions in 1st and 2nd order phase transitions at finite wave-vector is very different, in particular, the behavior of the second sound mode. We checked the finite $k$ results of QNMs for the 1st order phase transition, 0th order phase transitions as well as the ``cave of wind'' case, and get similar qualitative conclusions. Therefore we focus on the most representative results of 1st order phase transition in the rest of this section.

\begin{figure}[ht]
	\center
	\includegraphics[width=0.3\columnwidth]{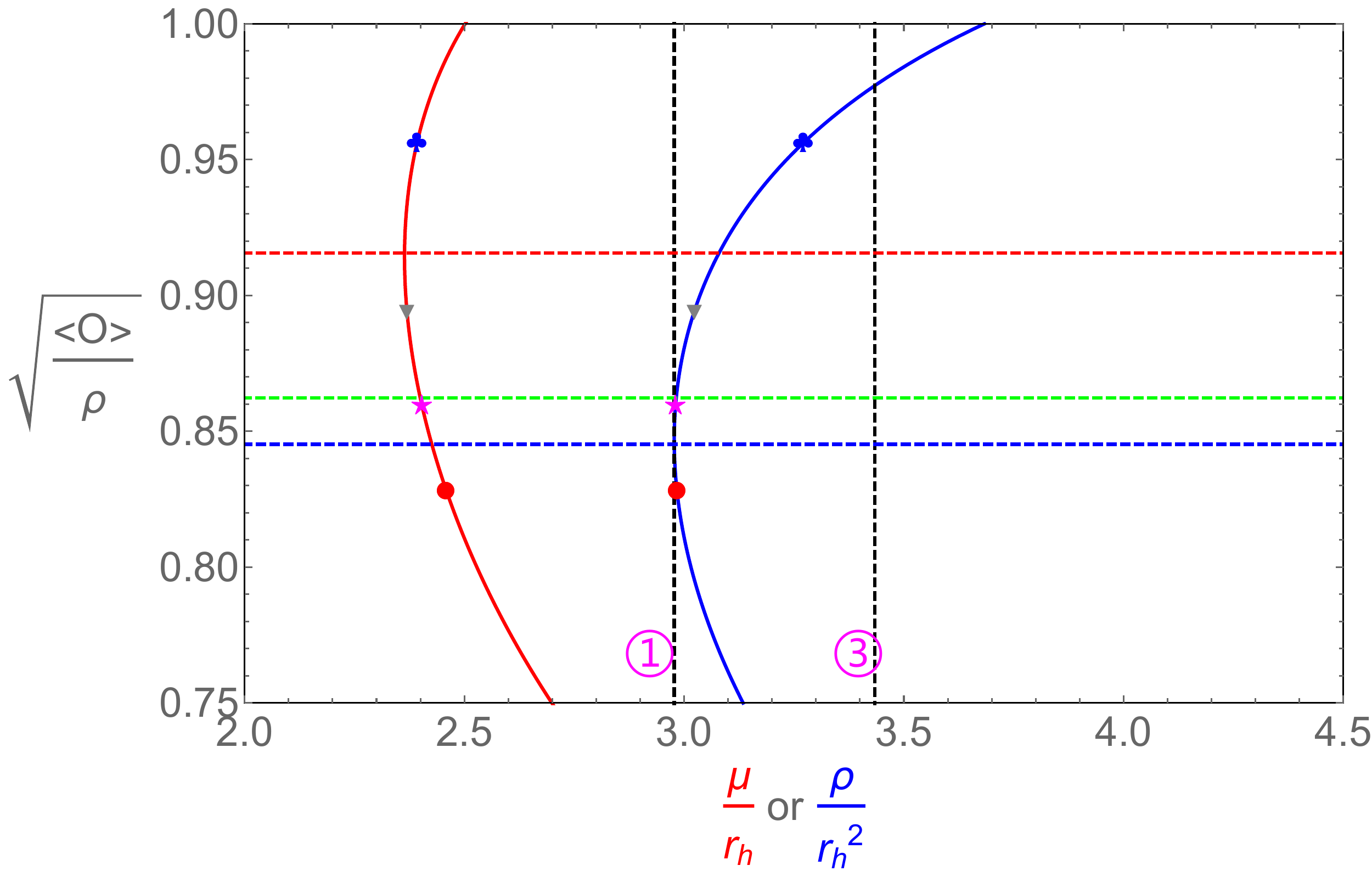}
	\includegraphics[width=0.3\columnwidth]{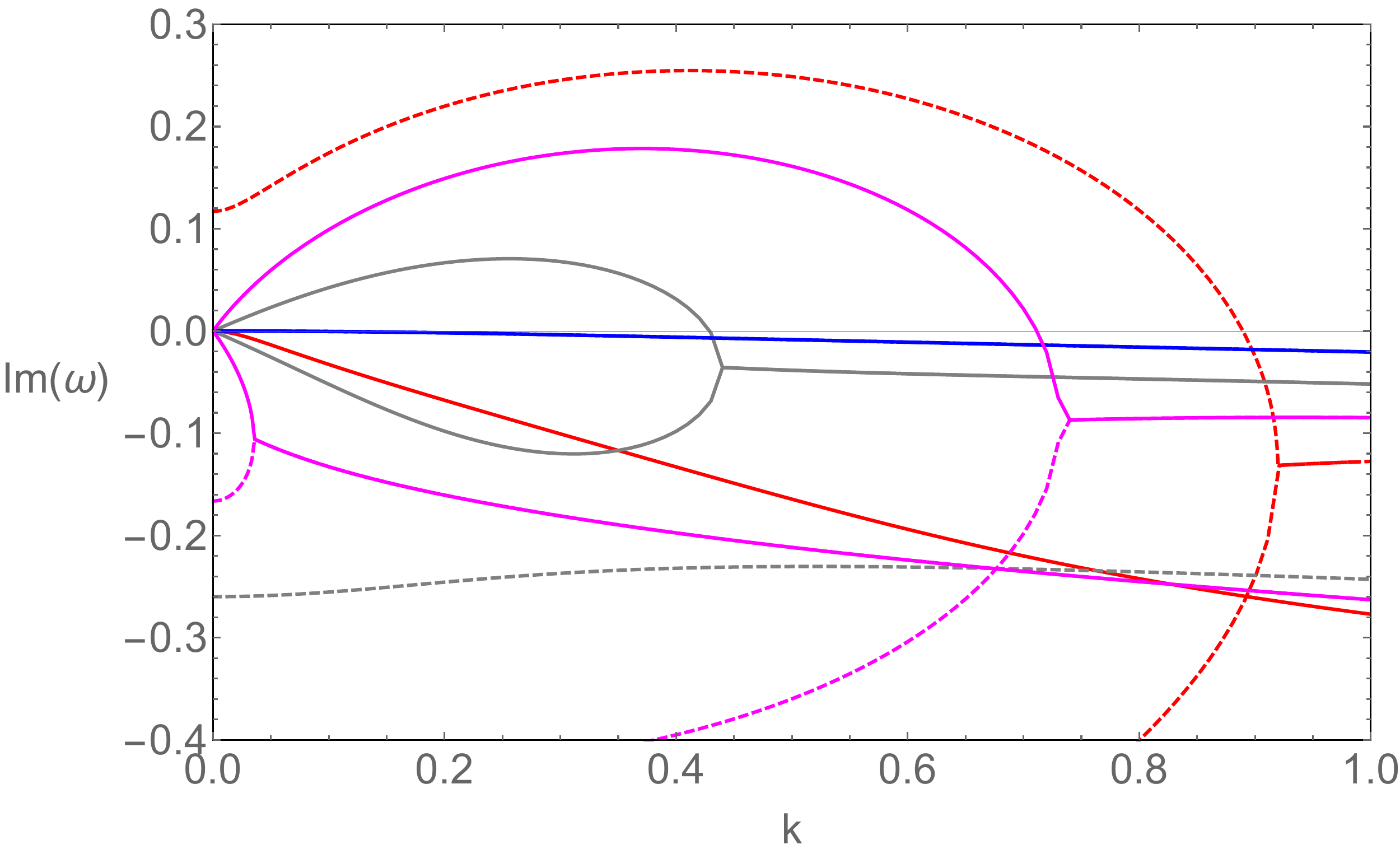}
	\includegraphics[width=0.3\columnwidth]{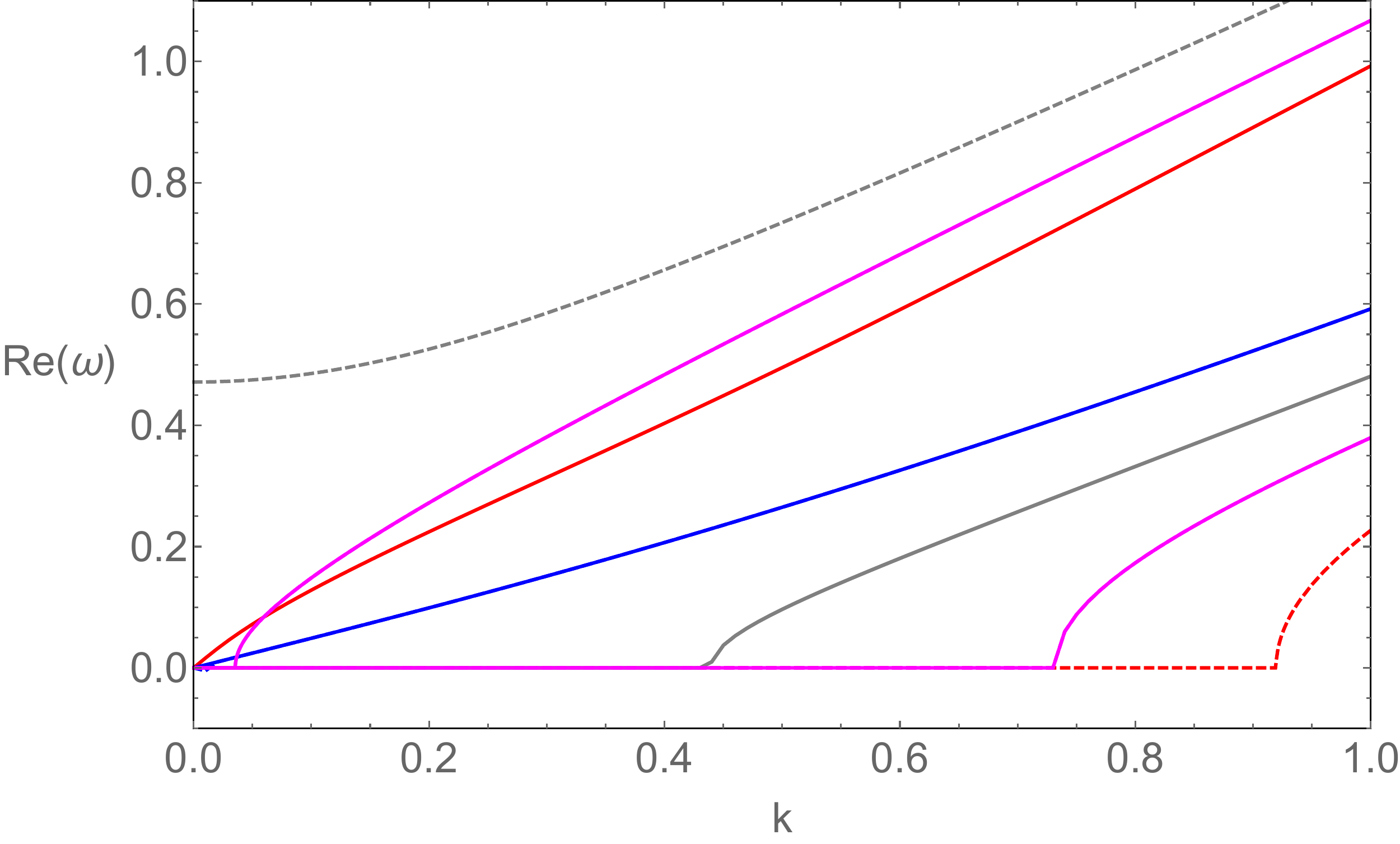}
	\caption{Quasinormal modes at finite wave-vector for the case of 1st order phase transition. The parameters of the potential are chosen as $\lambda=-2$ and $\tau=0.8$. \textbf{Left: }Blue and red solid lines are the condensate in the $\rho$ and $\mu$ ensemble respectively. The blue and red dashed lines indicate the location of the turning points. The green dashed line indicates the point where the two modes collide with each other as displayed in Fig.~\ref{condchange_omega_rho}. The two dashed black vertical lines marked by \ding{172} and \ding{174} show the positions for the turning point and the phase transition point, respectively, in the canonical ensemble as in Fig.\ref{condensate_diagram_landscape1st}. \textbf{Middle: }The imaginary part of the lowest QNMs. Each color corresponds to the points shown in the left panel and the different types of line to the different quasi normal modes. \textbf{Right: }the corresponding real part. Only the positive branch is shown. In both panels, the dashed lines indicate the amplitude mode and the solid ones the pair of sound modes.}\label{rho_mu_1st}
\end{figure}

In Fig.~\ref{rho_mu_1st}, we show the results for the 1st order phase transition with $\lambda=-2$ and $\tau=0.8$. The dependence of the imaginary part as well as the real part of the lowest lying QNMs on the wave-vector $k$ are plotted in the middle and right panels of Fig.~\ref{rho_mu_1st}, respectively. We use different colors to present results of different solutions, and mark the corresponding solutions on the condensate curve with the same color in the left panel of Fig.~\ref{rho_mu_1st}, where the two solid curves colored red and blue giving the relation of condensate with respect to the chemical potential $\mu/r_h$ and charge density $\rho/r_h^2$, respectively. For both the two curves, we use the same factor $\rho$ to get the dimensionless value of condensate as the vertical axes, therefore with the same value of vertical axes, one point on the red line present the same solution with another point on the blue line. Let us first focus on the blue line, which is the same to the condensate curve in the left panel of Fig.~\ref{condensate_free_1st}.

The Gibbs free energy and the resulting landscape picture for the homogeneous solutions show that the upper branch with larger value of condensate is (meta-)stable while the lower branch with smaller value of condensate is unstable, under homogeneous perturbations as confirmed by the $k=0$ results of QNMs. We mark this important turning point of the solid blue line by the dashed blue horizontal line in the left panel of Fig.~\ref{rho_mu_1st}. Then we see that the solutions marked blue, grey and magenta are on the upper branch while the red point is on the unstable lower branch.

With a first glance on the middle plot of Fig.~\ref{rho_mu_1st}, we can see that for the solution marked by the blue point, even with finite value of $k$, all the QNMs are in the lower half plane, indicating linear dynamical stability. However, for the solutions marked by the grey and magenta points, although the imaginary part of all the QNMs are not positive at $k=0$, a hydrodynamic mode moves upwards from the origin and get a positive value of imaginary part at small finite value of $k$. Although after reaching a maximum value, the imaginary part goes down and cross the horizontal axes at $k=k_0$, the positive imaginary part of this hydrodynamic mode at the region $0<k<k_0$ indicates instability under inhomogeneous perturbations. It is interesting that instability emerge in the upper branch which seems (meta-)stable in the landscape of homogeneous solutions. We can expect that the full landscape including all inhomogeneous configurations will explain the instability at finite wave-vector $k$. However even the full collection of all on shell inhomogeneous solutions are much more complicated, even impossible. Therefore it is interesting to study the origin of this special result from other perspectives. It is worth to notice that similar result is already discovered in Ref.~\cite{Janik:2016btb}, but in the spinodal region instead of the upper branch.

For the solution in the lower branch marked red, the amplitude mode showing instability at $k=0$ goes upwards at small $k$ and also go down across the horizontal axes after getting the maximum value. The lower branch is thus unstable under linear perturbation both at $k=0$ and finite $k$, although at this time the hydrodynamic mode presented by the solid red line in the middle and right panels of Fig.~\ref{rho_mu_1st} do not show any instability.

Because in our setup we leave the total charge conserved when perturbing the system to probe the linear dynamical stability, the system is in canonical ensemble. However, only in the case with $k=0$, the fixed value of total charge means the fixed value of local charge density $\rho$. Otherwise with finite $k$, the solution is perturbed in a non homogeneous way, therefore the charge of each subsystem in a local volume element is not fixed. The local conservation law implies that the charge only exchange between subsystems, and the total charge is still fixed. In this setup, supposing the state is still stable and the charge fluctuation finally dissipate. Then the chemical potential of each point is the same because there is no macroscopic charge flux. The subsystem is thus in equilibrium with neighbors with a fixed value of chemical potential, which is not canonical ensemble but grand canonical ensemble.

In order to test whether such analysis from the subsystem point of view explain the origin of the inhomogeneous instability in the upper branch in canonical ensemble, we include also the condensate curve in grand canonical ensemble as the solid red line in the left panel of Fig.~\ref{rho_mu_1st}, which show the condensate with respect to chemical potential $\mu/r_h$. This red curve also show a typical shape of 1st order phase transition. We also checked the grand potential with respect to the chemical potential, which also show a typical swallow tail shape, and the turning point is consistent with the turning point of the condensate curve. Therefore the landscape of homogeneous configurations at fixed value of chemical potential $\mu$ would show that the upper branch of the red solid condensate curve is stable while the lower branch of the red condensate curve is unstable under homogeneous perturbations with the value of chemical potential $\mu$ fixed. It is interesting to further study the QNMs under the condition of fixed value of $\mu$ in future studies.

It is obvious that the turning point of the solid red curve has a larger value of condensate $\langle \mathcal{O} \rangle_{t\mu}$ than the condensate of the turning point of the solid blue curve $\langle \mathcal{O} \rangle_{t\rho}$. Then the two special condensate value $\langle \mathcal{O} \rangle_{t\mu}$ and $\langle \mathcal{O} \rangle_{t\rho}$ divide the superfluid solution into three sections. With either fixed value of $\mu$ or fixed value of $\rho$, the section with $\langle \mathcal{O} \rangle>\langle \mathcal{O} \rangle_{t\mu}$ is stable while the section with $\langle \mathcal{O} \rangle<\langle \mathcal{O} \rangle_{t\rho}$ is unstable, under homogeneous perturbations. The section with $\langle \mathcal{O} \rangle_{t\rho}<\langle \mathcal{O} \rangle<\langle \mathcal{O} \rangle_{t\mu}$ is special, it is stable under homogeneous perturbations with fixed value of $\rho$ while it is unstable under homogeneous perturbations with fixed value of $\mu$. The grey and magenta points are both in this special region. We can see from the left plot of Fig.~\ref{rho_mu_1st} that this special region do not cross the 1st order phase transition point in the canonical ensemble marked by \ding{174}.
\begin{figure}
	\center
	\includegraphics[width=0.45\columnwidth]{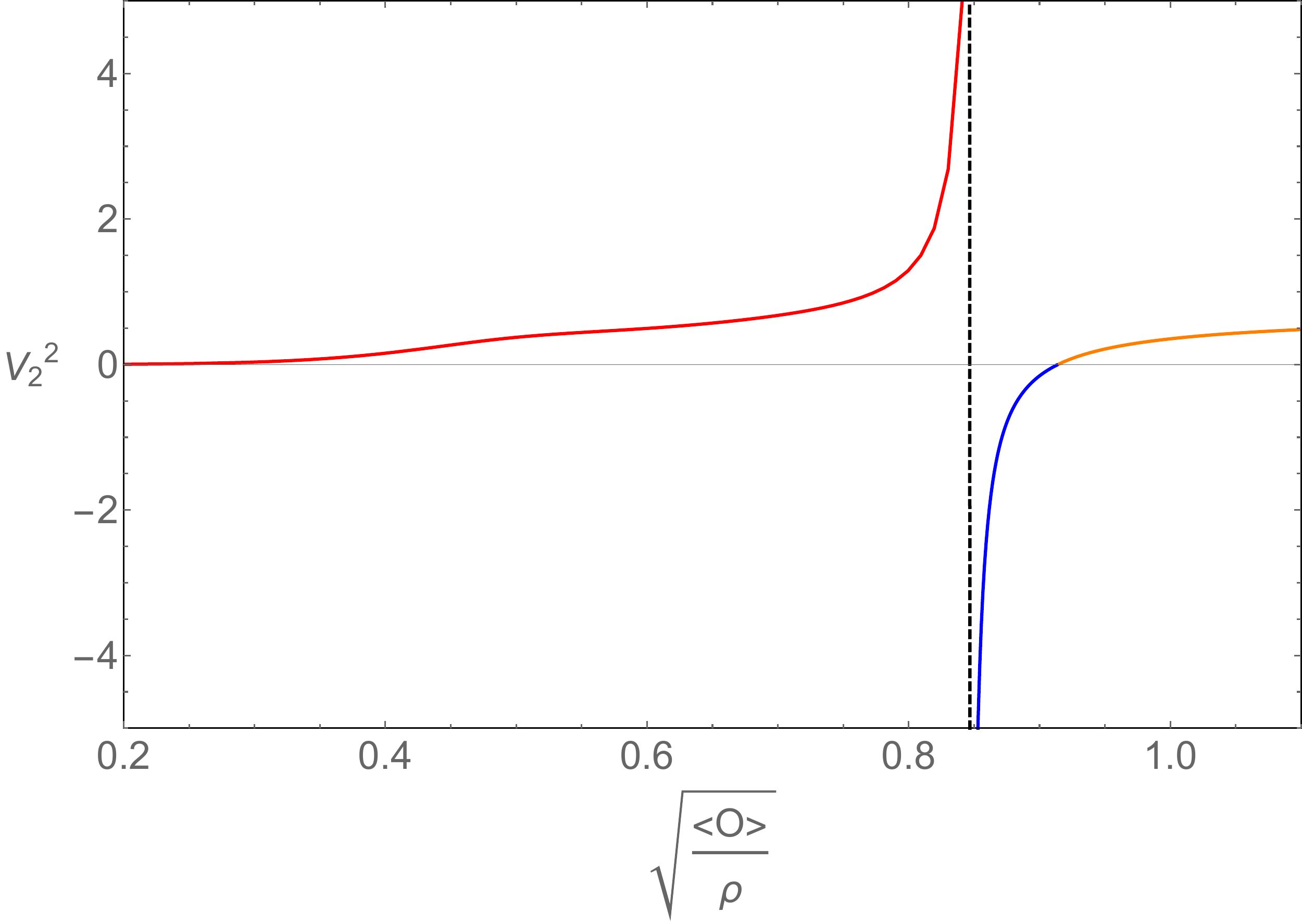}
    \includegraphics[width=0.45\columnwidth]{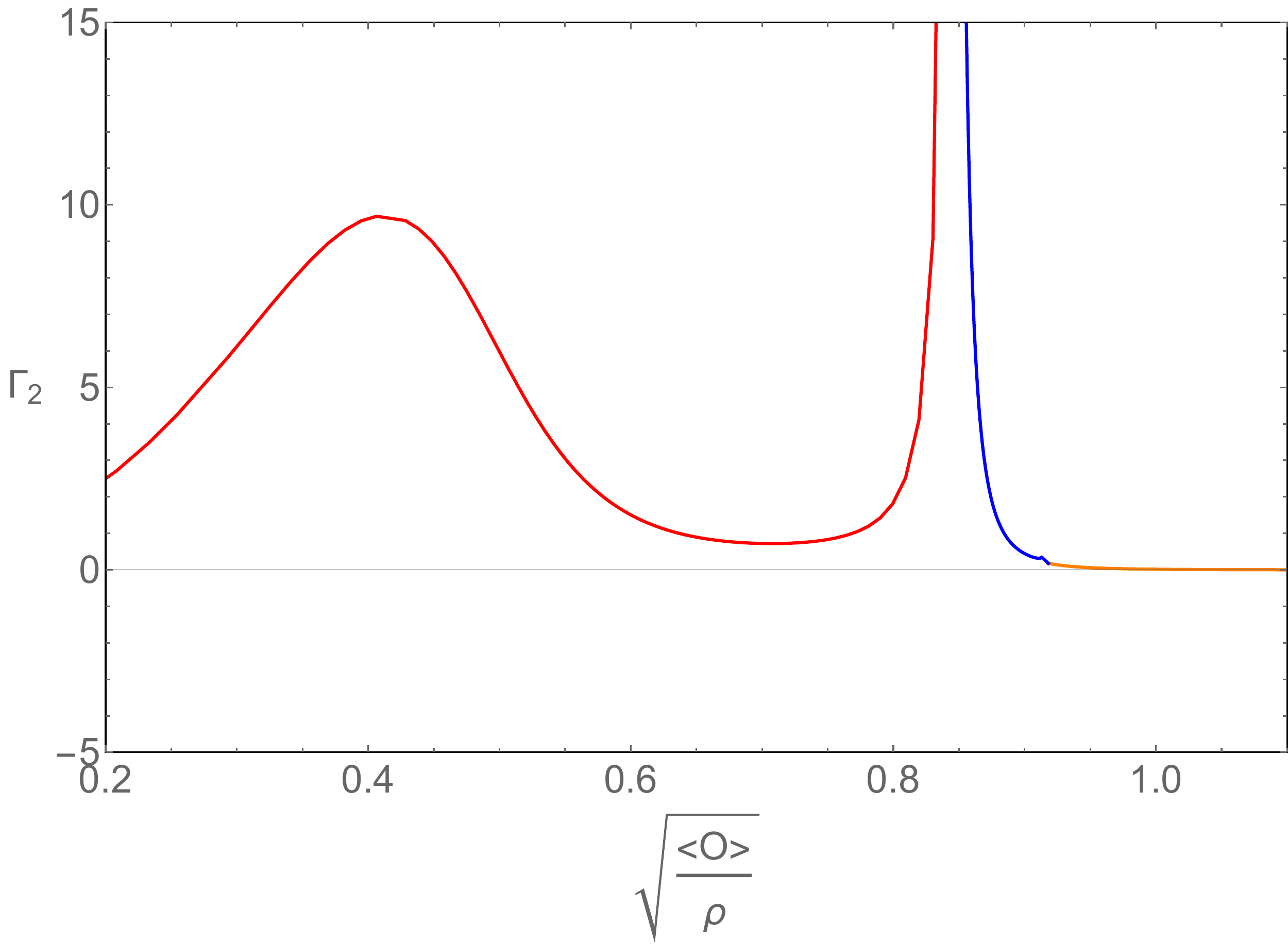}
	\caption{The speed and attenuation constant $v_2^2,\Gamma_2$ of second sound across the whole phase diagram in the case of the 1st order phase transition. The blue branch is the one in which the sound mode is unstable and it corresponds to the regime between the red and blue dashed lines in Fig.\ref{rho_mu_1st}.}\label{speed_of_sound}
\end{figure}

In order to confirm the region where the solution is stable under homogeneous perturbation but unstable under inhomogeneous perturbations, we plot the square of sound speed $v_2^2$ as well as the attenuation constant $\Gamma_2$ with respect to the condensate in Fig.~\ref{speed_of_sound}. In the left panel, we can see that the square of sound speed is positive both in the left region with small value of condensate and in the right section with large value of condensate. In the middle region the square of the sound speed is negative, indicating sound mode instability, and we confirmed that this region is exactly the section between the two turning points. We summarize our findings in Table \ref{tab1}.

\begin{table}[]
\centering
\begin{tabular}{| C{5cm} | C{2cm} | C{2cm}  |}
\hline
       regime      &  $k=0$    &    $k\neq 0$  \\
 \hline
 \hline
                  $\langle \mathcal{O} \rangle_{t\rho} < \langle \mathcal{O} \rangle_{t\mu} < \langle \mathcal{O} \rangle$          &     stable      & stable \\
                  \hline
                   $\langle \mathcal{O} \rangle_{t\rho} < \langle \mathcal{O} \rangle < \langle \mathcal{O} \rangle_{t\mu}$          &   stable         & unstable \\
                   \hline
                    $\langle \mathcal{O} \rangle < \langle \mathcal{O} \rangle_{t\rho} < \langle \mathcal{O} \rangle_{t\mu}$         &     unstable
         & {unstable}
         \\
 \hline
\end{tabular}
\caption{Summary of the results of the QNMs analysis for the 1st order phase transition. $\langle \mathcal{O} \rangle_{t\rho}$ and $\langle \mathcal{O} \rangle_{t\mu}$ are respectively the condensate values of turning points in the canonical and grand canonical ensembles (see the left panel of Fig.\ref{rho_mu_1st}). $k=0$ refers to homogeneous linear perturbations and $k\neq 0$ to inhomogeneous ones.}\label{tab1}
\end{table}

The full results of QNMs show that in the special region between the two turning points with $\langle \mathcal{O} \rangle_{t\rho}<\langle \mathcal{O} \rangle<\langle \mathcal{O} \rangle_{t\mu}$, the solution is stable under homogeneous perturbations but is unstable under inhomogeneous perturbations with fixed total charge. Since it is difficult to draw a picture of the full landscape including inhomogeneous configurations, or even get the full collection of on shell inhomogeneous solutions, we try to explain the origin of this inhomogeneous instability from a subsystem point of view as following.

Consider a subsystem in a small volume of the total system in canonical ensemble, when only homogeneous perturbations are considered, the charge density of the subsystem is also fixed. Therefore the stability is consistent with the landscape of the homogeneous configurations with fixed charge density. However, when non-homogeneous perturbations are considered in the form of QNMs with finite $k$, the condition of equilibrium of the local subsystem is the same value of chemical potential with neighbors.  Therefore the instability of the landscape for homogeneous configurations at fixed value of $\mu$ trigger the non-homogeneous instability of the total system. Since the section of negative square value of sound speed is the same to the section between the two turning points, this subsystem analysis perfectly explained the result of sound mode instability at finite $k$. We can also conclude that the QNMs of the gravity system correctly give the dynamical stability of the dual boundary field theory.

\begin{figure}[t]
	\center
	\includegraphics[width=0.6\columnwidth]{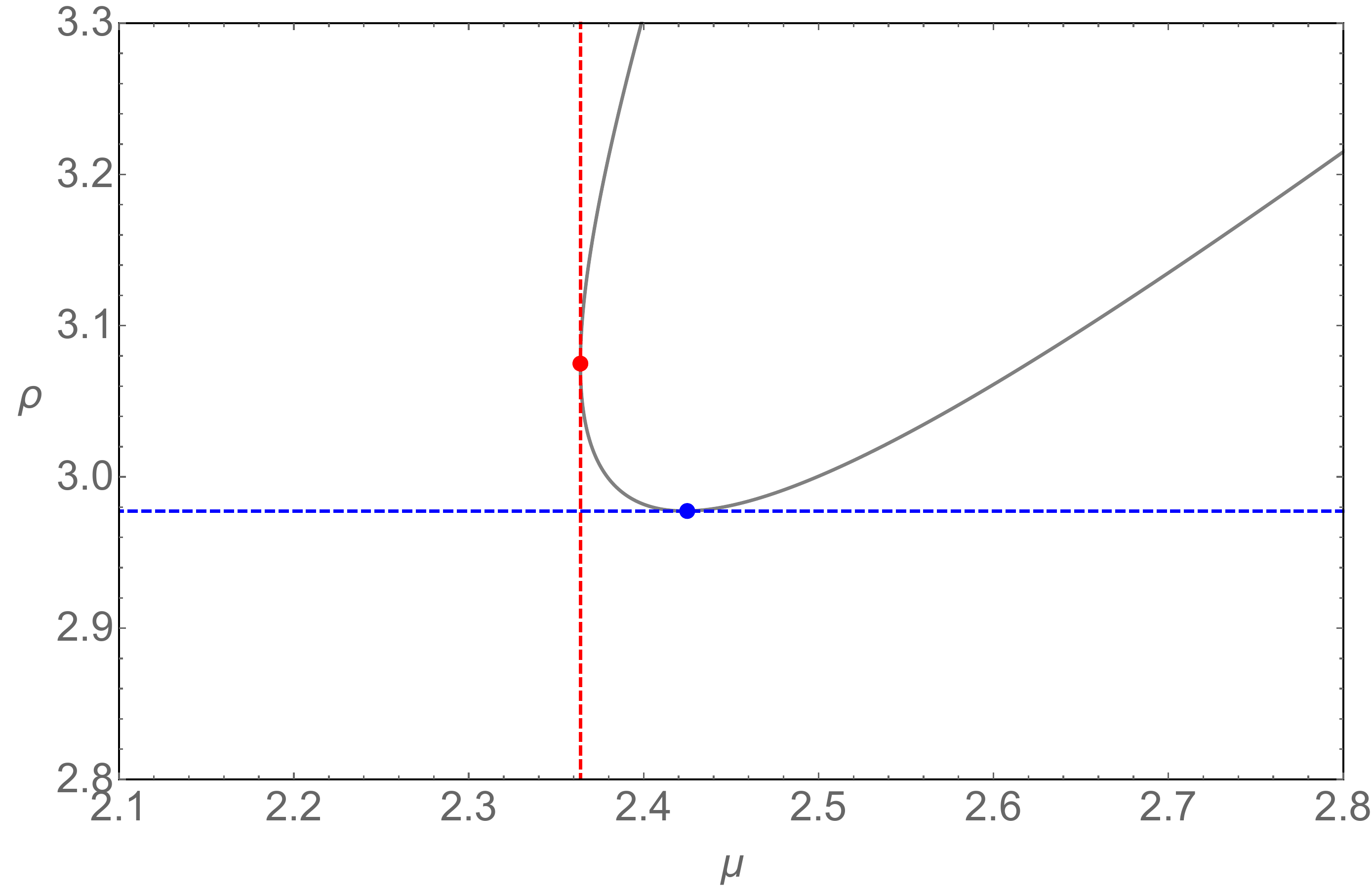}
	\caption{The relation between the charge density $\rho$ and the chemical potential $\mu$. The derivative of this curve coincides with the charge susceptibility $\partial\rho/\partial\mu$ and it is clearly negative between the blue and red dots. The latter corresponds to the turning points of the blue and red solid lines in the left panel of Fig.~\ref{rho_mu_1st}.}\label{relationship_rho_mu}
\end{figure}

The last issue is to check the relation of thermodynamic stability. The stability under homogeneous perturbations is well presented by the landscape picture of homogeneous configurations, which is speculated from the thermodynamic potential. The special region with sound mode instability is also consistent with the local thermodynamic instability signaled by the negative value of charge susceptibility $\partial\rho/\partial\mu$, which can be deduced by the positive value of $\partial \langle \mathcal{O} \rangle/\partial \rho$ and negative value of $\partial \langle \mathcal{O} \rangle/\partial \mu$ through
\begin{equation}
   \frac{\partial \langle \mathcal{O} \rangle}{\partial \mu}=
   \frac{\partial\rho}{\partial\mu}\frac{\partial \langle \mathcal{O} \rangle}{\partial \rho}.
\end{equation}
We plot the relation between the charge density $\rho$ and the chemical potential $\mu$ in Fig.~\ref{relationship_rho_mu} to show the section of negative susceptibility more concretely.

From the thermodynamic point of view, when the charge susceptibility is negative, once very small charge fluctuation exist, the subregion with more charge get lower value of chemical potential, while the neighbor with less charge get higher value of chemical potential. Therefore the charge flux directing from higher chemical potential to lower chemical potential further enlarges the fluctuation. This is the reason of the instability from inhomogeneous perturbations, and is consistent with our results of QNMs as well as analysis from landscape of homogeneous configurations in the two different ensembles. This consistency indicates a possible universal relation between dynamical instability and the local thermodynamical instability. We should emphasize that the dynamical instability is studied considering perturbations between on shell and off shell configurations with fixed value of some thermodynamic parameters, however, the local thermodynamical instability is studied considering the derivative between only on-shell solutions at different values of thermodynamical parameters.

Recall the similar discoveries in Ref.~\cite{Janik:2016btb}, where the boundary of the sound mode instability is not fully specified. With our new knowledge, we make a prediction that the other boundary of the sound mode instability is precisely the turning point of the condensate curve in the conjugate ensemble, and the susceptibility in this special region should be negative.

\section{Conclusions and outlooks}\label{sect:conclusion}
Let us summarize our rich discoveries in this simple holographic model.

1. We realize 1st and 0th order superfluid phase transitions as well as the ``cave of wind'' type phase transition in the simplest holographic superfluid model with two nonlinear terms in the probe limit. This model is a perfect choice for future studies involving the full time dependent dynamics, because the less number of field components involved will greatly simplify the numerics.

2. We discussed the landscape of homogeneous configurations of the different phase transitions, which is useful for better understanding the different phase transitions and predict the physics of dynamical stability of the system.

3. We calculated the quasi normal modes (QNMs) of the superfluid solutions. The $k=0$ results show consistency with the landscape of homogeneous configurations. Specially, the QNMs for the 0th order phase transition show that a standard 0th order phase transition can not occur and indicate global thermodynamic instability of the system. However, the system can be rescued to a ``cave of wind'' type, where the 0th order phase transition is replaced by a 1st order phase transition between superfluid solutions with different condensate.

4. The QNMs with finite wave-vector $k$ show that in a region where the system is (meta-)stable in the landscape of homogeneous configurations in canonical ensemble but is unstable in the homogeneous landscape in grand canonical ensemble, the system is stable under homogeneous perturbations, but unstable under inhomogeneous perturbations in a finite $k$ region. This is explained from the subsystem point of view as: when only homogeneous perturbation is considered, the local subsystem is also in canonical ensemble with charge density fixed, but when inhomogeneous perturbation is included, the local subsystem is in grand canonical ensemble instead. This region with only inhomogeneous instability is also consistent with the negative value of charge susceptibility, which indicates a probably universal relation between the inhomogeneous dynamical instability and the local thermodynamic instability.

There are several open questions which we would like to investigate in the near future. In particular, it would be interesting to explore the role of backreaction, the inhomogeneous dynamics, the time dependent evolution and the rich phenomena linked to the dynamics of phase transitions beyond the 2nd order type. We plan to explore some of these questions in the near future.

\section*{Acknowledgements}
We greatly thank Professor Matteo Baggioli for his valuable contribution to improve this work. We would like to thank Rong-Gen Cai, Xin Gao, Yu Tian, Li Li, Qi-Yuan Pan, Chuan-Yin Xia and Hua-Bi Zeng for useful discussions. We thank Li Li for useful comments on a first version of this draft.
This work is partially supported by NSFC with Grant No.11965013 and 11565017.


\bibliographystyle{JHEP}
\bibliography{reference.bib}
\end{document}